\newcommand{\be}{\begin{equation}}
\newcommand{\bea}{\begin{eqnarray}}
\newcommand{\ee}{\end{equation}}
\newcommand{\eea}{\end{eqnarray}}
\newcommand{\qa}{\alpha}
\newcommand{\qb}{\beta}
\newcommand{\qg}{\gamma}
\newcommand{\qG}{\Gamma}
\newcommand{\qd}{\delta}

\newcommand{\qe}{\varepsilon}

\newcommand{\qh}{\eta}

\newcommand{\qy}{\theta}

\newcommand{\qk}{\kappa}
\newcommand{\ql}{\lambda}
\newcommand{\qL}{\Lambda}

\newcommand{\qr}{\rho}
\newcommand{\qs}{\sigma}
\newcommand{\qS}{\Sigma}
\newcommand{\qt}{\tau}
\newcommand{\qf}{\varphi}
\newcommand{\qF}{\Phi}

\newcommand{\qj}{\psi}

\newcommand{\qo}{\omega}
\newcommand{\qO}{\Omega}

\newcommand{\sgn}{{\rm sgn}}

\newcommand{\tr}{{\rm tr}\;}
\newcommand{\Tr}{{\rm Tr}\;}

\newcommand{\inv}{^{-1}}

\newcommand{\dagg}{^{\dag}}
\newcommand{\prt}{\partial}

\newcommand{\intd}[1]{\int \!\! d{#1} \;}

\newcommand{\ointdx}{\oint\! dx\;}
\newcommand{\intdxx}{\int \!\! d^{2}x \;}

\newcommand{\fr}[2]{{\textstyle \frac{#1}{#2}}}
\newcommand{\half}{\mbox{$\textstyle \frac{1}{2}$}}

\newcommand{\naar}{\rightarrow}

\newcommand{\nn}{\nonumber}

\newcommand{\scez}{\setcounter{equation}{0}}
\newcommand{\ns}{\scez\section}
\renewcommand{\theequation}{\thesection .\arabic{equation}}

\newcommand{\pileft}{\stackrel{\scriptscriptstyle{\leftarrow}}{\pi}}
\newcommand{\nablaleft}{\stackrel{\scriptscriptstyle{\leftarrow}}{\nabla}}

\newcommand{\me}{m_{\rm e}}
\newcommand{\vd}{v_{\rm d}}
\newcommand{\veff}{v^{\rm eff}}
\newcommand{\eff}{^{\rm eff}}
\newcommand{\edge}{_{\rm edge}}
\newcommand{\Stop}{S_{\rm top}}
\newcommand{\tQ}{\widetilde{Q}}
\newcommand{\tI}{\tilde{I}}
\newcommand{\tIan}{\tilde{I}^\qa_n}
\newcommand{\Ian}{{\rm I}^\qa_n}
\newcommand{\Iamn}{{\rm I}^\qa_{-n}}

\newcommand{\curl}{\nabla\!\times\!}

\newcommand{\res}{^{\rm res}}

\documentstyle[preprint,aps,eqsecnum,epsf,prb,tighten]{revtex}
\begin{document}

\draft

\title{(Mis-)handling gauge invariance in the theory \\of the quantum Hall
 effect
III: \\ The instanton vacuum and chiral edge physics}

\author{A.M.M. Pruisken, B. \v{S}kori\'{c}, M.A. Baranov\cite{Misha}}

\address{Institute for Theoretical Physics, University of Amsterdam,
Valckenierstraat 65, \\ 1018 XE Amsterdam, The Netherlands}

\date{October 9, 1998}
\maketitle

\begin{abstract}
\noindent
The concepts of an instanton vacuum and ${\cal F}$-invariance are being
used to derive a complete effective theory of massless edge excitations in the
quantum Hall effect. Our theory includes the effects of disorder and Coulomb
interactions, as well as the coupling to electromagnetic fields and
statistical gauge fields. The results are obtained by studying the
strong coupling limit of a Finkelstein action, previously introduced
for the purpose of unifying both integral and fractional quantum Hall regimes.
We establish, for the first time, the fundamental relation between the
{\em instanton vacuum} approach and the completely equivalent theory
of {\em chiral edge bosons}. In this paper we limit the analysis to
the integral regime. We show that our complete theory of edge dynamics
can be used as an important tool to investigate longstanding problems
such as long-range, smooth disorder and Coulomb interaction effects.
We introduce a two dimensional network of chiral edge states and
tunneling centers (saddlepoints) as a model for smooth disorder. This
network is then used to derive a mean field theory of the conductances
and we work out the characteristic temperature ($T$) scale at which the
transport crosses over from mean field behavior at high $T$ to
the critical behavior plateau transitions at much lower $T$. 
The results explain the apparent lack of scaling which is usually seen
in the transport data taken from arbitrary samples at finite $T$.
Secondly, we address the problem of
electron tunneling into the quantum Hall
edge. We show that the tunneling density of states near the edge is
affected by the combined effects of the Coulomb interactions and 
the smooth disorder in the bulk. We express the problem in
terms of an effective Luttinger liquid with conductance parameter (g)
equal to the filling fraction ($\nu$) of the Landau band. Hence, even
in the integral regime our results for tunneling are completely
non-Fermi liquid like, in sharp contrast to the predictions of single
edge theories.  

\end{abstract}

\pacs{PACSnumbers 72.10.-d, 73.20.Dx, 73.40.Hm}

\tableofcontents

\ns{Introduction}
In problems of quantum
transport symmetries play an important role.
Recent advances in the theory of the quantum
Hall effect primarily make use of electrodynamic gauge invariance as
the fundamental symmetry of the strongly correlated electron gas.
\cite{bigWen,FrohlichZee} 
This symmetry permits one to proceed with a minimum of microscopic
input. Applications of Chern-Simons theory have been largely based upon
phenomenological arguments. These applications have provided a
universal language for the fractional quantum Hall effect in which the
various hierarchy schemes could be treated on equal
footing. \cite{bigWen}

Application of Chern-Simons theory has also led to the idea that many of
the basic properties of incompressible quantum Hall states can be
understood in terms of Luttinger liquid behavior of the edge
excitations. 
This non-Fermi liquid theory of edge excitations is now commonly used as
a computational scheme for tunneling properties of different quantum
Hall states as well as the thermodynamic properties of the fractionally
charged quasiparticles.
It is important to keep in mind, however, that unlike the
conductance parameters, physical quantities like the tunneling density
of states do not necessarily follow the rules of incompressibility.
The lack of a microscopic theory of the fractional
quantum Hall effect has led to controversial issues regarding
the {\em definition} of the Hall conductance (notably for those states
that have edge channels of opposite chirality).
\cite{KaneFisher,Haldane} Moreover,
serious discrepancies have arisen between the predictions of the Luttinger
liquid theory of edge excitations\cite{KaneFisher} on the one hand and
the experimental
results on edge tunneling on the other.\cite{continuum}

This paper is the third in a series of articles in which we lay
down the foundation
for a microscopic theory of disordered compressible and incompressible
states in the (fractional) quantum Hall regime. In previous
papers\cite{MishandlingI,MishandlingII}
(hereafter called I and II) we introduced an effective Finkelstein
action for localization and interaction effects. The Finkelstein
action includes the
topological concept of an instanton vacuum as well as the statistical
(Chern-Simons) gauge fields. The inclusion of statistical gauge fields
in the problem makes it possible to formulate a combined theory of
composite fermions, localization and interaction effects.
The results of {\em weak coupling} analyses (both perturbative and
non-perturbative, i.e. instantons) can then be used
to obtain a global scaling diagram for the conductances. 
The integral as well as the fractional quantum Hall regime are
incorporated in this scaling diagram.
In this work, we are primarily interested in the strong coupling limit
of our action where
the system has a gap in the density of states.
This physical situation is the same as the one described by the 
Chern-Simons approach with one important exception: beside the Coulomb
interactions we also deal from first principles with the effects of
disorder. 

\vskip5mm

\noindent
One of the main objectives of this work is to derive microscopically a
Luttinger liquid theory for edge excitations in the presence of disorder
and electron-electron interactions. From our general, effective action
point of view we can say that the physics of edge excitations has
fundamental significance since it provides unique and invaluable
information on the topological concept of an `instanton vacuum'\cite{Pr} in
strong coupling.

An additional important advancement is that we 
obtain for the first time the {\em
complete} Luttinger liquid theory on the edge. We have the action for
interacting
chiral edge bosons coupled to external electromagnetic fields.
This theory can now be used to define the Hall conductance in a
general, unambiguous manner by expressing the appearance of an
`edge anomaly' \cite{Balachandran} in terms of Laughlin's gauge
argument.\cite{Prange}

The details of the analysis of edge excitations are described in
sections \ref{secderivfull} and \ref{secchiral}. This analysis is
based, to a large
extent, on the various new concepts which were introduced in [I] under the
name of `${\cal F}$-algebra' and `${\cal F}$-invariance'. Recall that
in [II] we also studied these concepts but in the weak coupling regime.
This paper therefore shows that ${\cal F}$-invariance
retains its significance all the way down to the regime of strong
coupling, where the massless excitations are confined to the edges of
the sample.
It is important to note that this is the first time that this symmetry
is being demonstrated in the weak as well as the strong coupling regime.

The results of sections \ref{secderivfull} and \ref{secchiral} will
serve as the starting point for a microscopic theory of edge excitations
in the fractional quantum Hall effect. We shall limit ourselves here to
the integer regime, since this already contains most of the difficulties. 
Extensions of our theory to include the fractional effect can be done by
means of the statistical gauge fields. These will be reported elsewhere.

We shall begin
by reviewing and extending the topological instanton vacuum approach
to the qHe, following the ordinary, free electron replica formalism
in section II.
In making the connection between topology and edge currents, we show
that important aspects of the problem have previously been
overlooked. In particular, we show that the {\em massless} excitations
of the disordered edge states are obtained from the {\em fluctuations}
about {\em integer} quantized {\em topological charge} (section A3).
This important observation will serve as a starting point for most of the
analyses in the remainder of this paper.

{\em Massless} edge excitations appear in the instanton vacuum theory
for arbitrary number of field components (replicas) $N_r$ and not just
in the replica limit $N_r\!=\!0$. The present analysis revises our
previously accumulated knowledge of the subject in at least two respects.
First we recognize that a direct
relationship exists between the numerical value of the instanton
parameter $\theta$ (or $\qs_{xy}^0$, Ref.~11) and the
phenomenon of 
{\em inter-channel scattering} at the edge. 
Here the number of edge channels equals the number of fully occupied
Landau levels, and the phrase ``inter-channel scattering'' refers to
the effect of a random short-ranged potential.

Secondly, we review the
earlier attempts toward establishing a general {\em topological
principle} for quantization of the Hall conductance which includes the
effect of localization of the bulk states. 
The mere existence of massless edge excitations turns out to have
basic consequences for the {\em quantization phenomenon} which now can
be shown to be a robust and fundamental aspect of the instanton vacuum
theory with arbitrary values of $N_r$.

\vskip5mm

\noindent 
In all our work sofar, we have substituted the phrase `electronic
disorder' for a white noise random potential.
This was always done for technical reasons alone. 
However,it is well known that in real quantum
Hall devices slowly varying potentials are often present.
\cite{Prange,PrW}
Till now
these have been in general difficult to handle. Our microscopic theory
of the edge enables us to treat long-range potentials as well as
electron-electron interactions. In this paper we embark on solving
two longstanding problems where smooth disorder and Coulomb
interactions give rise to unexpected results. By addressing these
problems we attack the core of the controversies that exist between
the theory and experiments that presently span this subject.

The first problem we address is that of the plateau transitions.
This we model as a percolating network of `edge states'
(equipotential contours) and widely separated `saddlepoints'. 
A large class of such systems is then `mapped' onto the non-linear
$\qs$ model representation for localization, and the main problem is
to identify the length and energy scales of the `bare' parameters, or
the mean field conductances which together determine the
renormalization starting point, i.e. the point where scaling occurs
first.
This starting point can involve, in principle, arbitrarily large
distances and arbitrarily small energies and this, obviously,
complicates the observability of the critical behavior of the
Anderson (plateau) transitions.
We argue that Coulomb interaction effects lead to a modified mean
field theory of transport which is now observed in the experiments
performed at finite temperatures.
The chiral boson theory shall be used to actually compute the
inelastic relaxation rate of the conducting electrons in the
saddlepoint network.
This, then, might conceivably be the explanation for the empirical
fits of the transport data taken recently from presently available
samples. \cite{semiclass}

As the second typical example of long-ranged disorder effects we
embark on the problem of electron tunneling into the quantum Hall
edge. We show that the Coulomb interactions between the edge and the
`localized' bulk orbits dramatically differ from the predictions of theories
which are based on isolated edges alone.
Tunneling processes into the quantum Hall edge have, in fact, nothing
to do with the quantization of the Hall conductance or the
`incompressibility' statement which describe the non-equilibrium
properties of the electron gas. 
We find that the tunneling density of states near the edge can be
understood in terms of an effective edge theory
which describes the equilibrium properties of the combined
edge and bulk degrees of freedom.
The Luttinger liquid parameter $g$ is related to the filling fraction
$\nu$ of the bulk Landau level. This leads to a tunneling exponent
which varies like $1/\nu$, in agreement with recent experimental data
on the tunneling current, taken from samples in the fractional quantum
Hall regime.\cite{continuum}
This situation is dramatically different from what is expected while
assuming an isolated edge, or in the case of short-ranged disorder
which gives rise to scattering between different edge states.\cite{KaneFisher}

In this paper and one that follows\cite{fracedge} we carefully re-examine the
consequences of inter-channel edge scattering. We reproduce the
completely different Kane-Fisher-Polchinsky\cite{KaneFisher} 
scenario of tunneling
exponents in the integral and fractional regimes from our strong
coupling edge theory.
However, we argue that both the assumptions (an isolated edge and
short-ranged disorder or inter-channel scattering) are clearly
incorrect since the problem is two dimensional and dominated by 
long-ranged potential fluctuations as well as interaction effects.

\vskip5mm

\noindent
The organization of this paper is as follows.

In Section~\ref{secedgeexc} 
we introduce the problem in the language of the replica
free electron theory. We briefly recall the instanton vacuum approach
in (A1).\cite{Pr} The connection between topology and interchannel scattering
between the chiral edge modes is made in (A2). 
This leads to an exact solution of the instanton vacuum at the edge
which can now be shown to be critical (A3).

In Section~\ref{secplatrev} 
we introduce a two dimensional network of chiral edge
states as a model for the problem of long-ranged potential
fluctuations. This is then used for mean field purposes and for
demonstrating universality of the plateau transitions in (B1). In (B2)
we extend the network approach to include interaction effects. A
semiclassical theory of transport is introduced in order to
explain the lack of scaling recently found 
in many (ordinary) quantum Hall devices at
finite temperatures. (B3) contains several general remarks.

In Section~\ref{secderivfull} 
we present a detailed derivation of the
complete chiral edge theory using the fermionic path integral.  

In Section~\ref{secchiral} 
we make the fundamental connection between the {\em
instanton vacuum} on the one hand and {\em Chern Simons gauge theory} and
{\em chiral edge bosons} on the other. 

In Section~\ref{seclongrange} 
we apply the theory of chiral edge bosons to several
problems of long-range disorder and interaction in the bulk of the
sample. These include the density of states for tunneling into the quantum Hall
edge as well as the relaxation times entering into the transport
problem of Section \ref{secplatrev}.

We end this paper with a summary in Section~\ref{secsummary}.

\ns{Edge excitations}
\label{secedgeexc}
\subsection{Sigma model}
\label{secsigmamodel}

\begin{figure}
\begin{center}
\setlength{\unitlength}{1mm}
\begin{picture}(140,80)(0,0)
\put(0,0)
{\epsfxsize=140mm{\epsffile{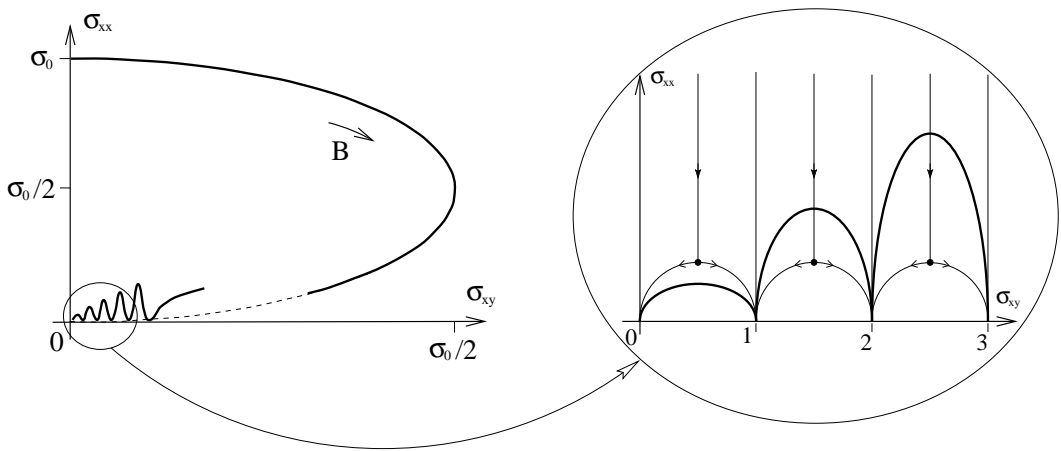}}}
\end{picture}
\caption{Sketch of the mean field conductances for a short-range
disorder potential. The inset is the strong field limit or
quantum Hall regime. The renormalization group flow
lines indicate how the mean field theory results change after
successive length scale transformations. After Refs.~8, 11 and 15.}
\label{figzoomsigma}
\end{center}
\end{figure}

Let us recall the instanton vacuum theory \cite{Pr,PrinPrange}
for the integral quantum Hall
effect which is expressed in terms of the local field variables
$Q^{\qa\qb}_{pp'}$, where $\qa,\qb\! =\! 1,\ldots,N_r$ are the replica indices
and $p,p' \! =\!\pm 1$ are the indices denoting advanced/retarded waves. They
can be represented as
\bea
	Q=T\inv\qL T \hskip5mm & {\rm with } &\hskip5mm \qL^{\qa\qb}_{pp'}=
	\qd^{\qa\qb}\qd_{pp'}\sgn(p)
\label{defQ22}
\eea
and $T$ a unitary matrix of size $2N_r\!\times\! 2N_r$. The complete
action is given by
\be
	S[Q] = -\fr{1}{8}\qs^0_{xx}\intdxx \tr(\nabla Q)^2
	+\fr{1}{8}\qs^0_{xy}\intdxx \tr\qe_{ij}Q\prt_i Q\prt_j Q
	+\pi\qr_0\qo \intdxx \tr \qL Q.
\label{Ssigma}
\ee
Here $\qs^0_{ij}$ stands for the mean field conductances in units of
$e^2/h$ (see Fig.~\ref{figzoomsigma}), 
$\qr_0$ is the (exact) density of states at the Fermi energy
and $\qo$ is the frequency. 
The second term in (\ref{Ssigma}), proportional to the mean field Hall
conductance ($\qs_{xy}^0$), has remained one of the most difficult
chapters in the theory of Anderson localization in low dimensions.
Most of the insight into the theory with $N_r\!=\!0$ number of field
components has come from weak
coupling renormalization theory (both perturbative and
non-perturbative, i.e. instantons). \cite{Pr}
In particular we mention the
global scaling diagram of the conductances as well as the appearance
of a critical fixed point in strong coupling regime.\cite{PrinPrange} 
This fixed point theory predicts a massless (metallic) phase
at the Landau band center as well as the following scaling result for
the conductances\cite{Prl}
\bea
\sigma_{ij} (L,B)= g_{ij} ([L/\xi]^{1/\nu} )
	& \hskip5mm ; \hskip5mm & 
	\xi = |B-B^*|^{-\nu}
\label{condscaling}
\eea

\noindent
which cannot be obtained in any different way.
Here, the function
$g_{ij} (X)$ is a regular function of its argument, $B^*$ is the
critical magnetic field strength and $\nu$ stand for the critical
index for the localization length $\xi$.
Following the experimental tests of (\ref{condscaling}) 
by H.P. Wei {\it et al.} \cite{HPWei}   
extensive numerical work on the free electron gas has been performed
and the quoted best value for the critical index is $\nu=2.3$.\cite{Kramer}

To date, no exact (conformal) scheme for the critical indices
exists. All that one can say at this time is that the field of exactly
solvable models is not sufficiently developed to be able to handle the
specific subtleties of topology and replica field theory.
These subtleties are all well
understood within the elaborate framework of weak coupling expansion
techniques\cite{Pr}
and the results were used to unfold and predict the entire 
singularity structure of the theory, notably (\ref{condscaling}). 

In previous work \cite{PrB} we have shown that the 
theories of free and interacting electrons share the same basic
features such as asymptotic freedom, instantons etc. The same scaling
diagram for the conductances was obtained, which means that
(\ref{condscaling})
remains valid also when the Coulomb interactions are taken into
account. This important result was conjectured but otherwise
completely un-understood at the time of the original experiments on
criticality.

\subsubsection{Strong coupling}
\label{secstrongcoupl}
In this paper we address the subtleties of the instanton vacuum theory in an
extremely important exactly solvable limit where
$\rho_0 \!=\!\qs_{xx}^0\! =\! 0$
and where the Hall conductance is integer quantized
($\qs_{xy}\!=\!m$). Physically this happens when the Fermi energy is
located in a density of states gap between adjacent Landau bands. 
In this {\em strong coupling} limit massless excitations do exist at
the edges of the system.
Since several, basic aspects of the problem have previously gone
unnoticed we shall proceed first within the free electron formalism
of Eqs.~(\ref{defQ22}) and (\ref{Ssigma}). 
We come back to the fermionic path integral in
Sections \ref{secderivfull} and \ref{secchiral}.   

For $m$ completely filled Landau levels the action becomes
simply
\be
	S[Q]=\fr{m}{8}\intdxx\tr\qe_{ij}Q\prt_i Q\prt_j Q
	= \fr{m}{2}\oint\! d\vec x\cdot \tr(\qL T\nabla T\inv)
\label{Sgap}
\ee
where the surface integral is taken over the sample's edge.
Recall that (\ref{Sgap}) is quantized according to
\be
	S[Q]=2\pi im\cdot q[Q]
\label{quantq}
\ee
with $q$ the integer topological charge, provided that the $T$-matrix
reduces to a $U(N_r)\!\times\! U(N_r)$ gauge at the edge.\cite{PrinPrange}

Under these circumstances the sample edge has been contracted to a
single point (spherical boundary conditions)
and (\ref{quantq}) is a realization of the formal
homotopy theory result 
$\pi_2(G/H)\! =\! Z$ which states that the mapping of
$Q$ onto the 2D plane is described by a set of integers $q$. It is
natural to take the theory one step further and propose the
quantization of the charge $q[Q]$ as the topological principle in
replica field theory which forces the Hall conductance ($m$) itself to
be integer quantized. The idea has led to a consistent quantum theory of
conductances that unifies a fundamental aspect of asymptotically free
field theory (i.e. dynamic mass generation) with the quantum Hall
effect.\cite{Pr} 
More specifically, it says that the conductances in (\ref{condscaling})
always scale toward $\sigma_{xx}\!=\!0, \sigma_{xy}\!=\!m$ 
for $L$ large enough.

One can show \cite{PrinPrange} that the $U(N_r)\!\times\! U(N_r)$
gauge condition at the edge is the replica field theory version of a
static $U(1)$ gauge acting on the physical edge states. Such a $U(1)$
gauge implies that an integer number of edge levels has crossed the
Fermi level. This level-crossing is necessarily induced by the averaging
procedure over random potentials.

Nevertheless, it is somewhat disappointing to know 
that the topological invariant
in (\ref{Ssigma}), as it was discovered originally in a microscopic 
derivation, is truly defined with {\em free}
boundary conditions and without any separation between edge and bulk
degrees of freedom.\cite{Pr} Sofar, the precise significance of
boundary conditions has remained obscure. 

\subsubsection{Interchannel edge scattering}
\label{secHmatrix}
In what follows, we show that the fluctuations about precisely
quantized values for the topological charge represent, in fact,
essential physics of the problem, since they describe the dynamics of
(massless) edge excitations. In order to see this, we write $T$ as the product
of a $U(N_r)\!\times\! U(N_r)$ gauge $U$ and a small fluctuation $t$,
\be
	T=Ut.
\label{Tut}
\ee
The action now becomes
\be
	S[Q]=2\pi im\cdot q[U]
	+\fr{m}{2}\oint\! d\vec x\cdot \tr(\qL t\nabla t\inv)
	+\pi\qr\edge\qo\oint\! dx\; \tr\qL Q
\label{Ssplit}
\ee
with $\qr\edge$ the density of edge states. 
One way of identifying (\ref{Ssplit}) as the effective theory of
disordered chiral edge states is to redo the derivation, but now for
the 1D system with Hamiltonian
\be
	{\cal H}\edge=-i\vd\prt_x+V(x)
\label{1Daction}
\ee
where $\vd$ is the drift velocity of the edge electrons and $V(x)$ the
random potential. 
It turns out that our initial guess (\ref{1Daction}) is correct only
in the case 
$m\! =\! 1$ in (\ref{Ssplit}). This problem is easily resolved
once one realizes that $m$ really stands for the number of filled
Landau levels, such that (\ref{1Daction}) should be replaced by a
Hamiltonian for a total of $m$ edge channels. Hence, an obvious second
guess would be
\be
	{\cal H}\edge=\sum_{j=1}^m {\cal H}\edge^{(j)}
\ee
where ${\cal H}\edge^{(j)}$ is the same for all $j$, i.e. each of the
$m$ eigenstates experiences the same white noise potential $V(x)$,
just as it appears in the original problem in two spatial dimensions.
This, however, is not correct and the theory with general $m$, 
(\ref{Ssplit}), necessarily requires inter-channel scattering to take
place. We have to start from a matrix Hamiltonian
\be
	{\cal H}\edge^{jj'}=-i\vd\qd_{jj'}\prt_x
	+V_{jj'}(x)
\label{Hjj}
\ee
where $V$ is a Hermitian matrix. The matrix elements $V_{jj'}$ connect
the edge channels $j,j'$ and are distributed with a weight
\be
	P[V]\propto \exp\{-\fr{1}{g}\ointdx
	\tr V^2\}.
\ee
One can construct a generating function for the free particle Green's
functions as usual, according to
\be
	Z=\int\!\!{\cal D}[\bar{\qj}\qj]\int\!\!{\cal D}[V] P[V]
	\exp \sum_{p=\pm,\qa,jj'}\ointdx \bar{\qj}_p^{\qa,j}
	\left[(\mu+ip\qo)\qd_{jj'}-{\cal H}\edge^{jj'}
	\right]\qj_p^{\qa,j'}.
\label{multipsi}
\ee
In Appendix C we show that (\ref{multipsi}) and (\ref{Ssplit})
are identical in the limit of large distances.

\subsubsection{Criticality at the edge}
We next point out that the results of the previous section provide an
exact solution to our topological theory at the edge
(\ref{Sgap}--\ref{Ssplit}) for all values of $N_r$. The simple but
important observation to be made is that the random potential
$V_{jj'}(x)$ in (\ref{multipsi}) can be `gauged away', i.e. absorbed in
a redefinition of the fermion fields, and all that remains is the
trivial theory of `pure' chiral edge states,
\be
	Z=\int\!{\cal D}[\bar\qj \qj]\exp\sum_{p=\pm,\qa,j}\ointdx
	\bar\qj^{\qa,j}_p[-i\vd\prt_x+i\qo p]\qj^{\qa,j}_p.
\label{purestates}
\ee
Eq.~(\ref{purestates}) is just a formal way of saying that edge
electrons do not Anderson localize, because chirality excludes
backscattering processes on random impurities. 
Following up on the analysis of Appendix~C we will use the simplicity
of (\ref{purestates}) and derive explicit 
results for the $Q$-field (\ref{Ssplit}).
Write
\bea
	G_\pm^{jj'}(x,x')&=&\langle x,j|(-{\cal H}_{\rm edge}\pm
	i\qo)\inv |x',j'\rangle  \\
	G_\pm(x,x')&=&\langle x|(i\vd\prt_x\pm i\qo)\inv|x'\rangle
\eea
to represent the single particle propagator of the dirty edge
(\ref{multipsi}) and clean edge (\ref{purestates}) respectively.
Some useful identities are given by,
\be
	\qr_{\rm edge}=\fr{1}{2\pi i}\sum_j\left[G^{jj}_-(x,x)
	-G^{jj}_+(x,x)\right]
	=\fr{m}{2\pi i}\left[G_-(x,x)-G_+(x,x)\right].
\ee
Here $\qr_{\rm edge}$ denotes the density of edge states at the Fermi
level which can be obtained explicitly from the r.h.s.,
\be
	\qr_{\rm edge}(x)=\qr_{\rm edge}=m/2\pi\vd.
\label{defre}
\ee
Eq.~(\ref{defre}) shows that the density of edge electrons is a
constant, independent of $x$ and disorder, as it should be.
An important conclusion now follows for the theory of $Q$-fields
(\ref{Ssplit}), namely
\be
	\langle Q\rangle = \qL
\label{QisqL}
\ee
(where the expectation is with respect to (\ref{Ssplit})),
which holds for arbitrary $N_r$. This result may be obtained e.g. by
differentiating both theories (\ref{Ssplit}) and (\ref{multipsi},
\ref{purestates}) with respect to $\qo$.
Notice that (\ref{QisqL}) can be regarded as the `order parameter'
(analogous to the magnetization in the language of the Heisenberg
ferromagnet) and one would naively expect this quantity to vanish in
one spatial dimension. The result $\langle Q\rangle = \qL$ indicates,
however, that the continuous symmetry is permanently broken at the
edge of the instanton vacuum for all numbers of field components
$N_r$. This apparent violation of the Mermin-Wagner-Coleman theorem is
clearly due to the lack of positive definite Boltzmann weights in our
problem that is described by an imaginary action (\ref{Ssplit}).
Eq.~(\ref{QisqL}) also indicates that the edge of the topological
vacuum is critical. The simplest way of demonstrating this is by
employing the background field method.
For example, the replacement $t\naar t \cdot t_0$ in the second term
of (\ref{Ssplit}) can be written as
\bea
	\ointdx \tr[\qL t\prt_x t\inv] &\naar &
	\ointdx \tr[\qL tt_0\prt_x (t_0\inv t\inv)] \nn\\
	&=&
	\ointdx \tr[\qL t\prt_x t\inv]
	+\ointdx \tr[Q t_0\prt_x t_0\inv].
\eea
Here, $t_0$ represents a fixed and slowly varying background field. We
obtain an effective action for $t_0$ as follows,
\be
	S_{\rm eff}[t_0]=\fr{m}{2}\ointdx 
	\tr[t_0\prt_x t_0\inv\langle Q\rangle]
	=\fr{m'}{2}\ointdx \tr[\qL t_0\prt_x t_0\inv].
\label{background}
\ee
Eq.~(\ref{background}) defines an `effective' parameter 
$m'\!=\!m \;\tr\qL\langle Q\rangle/2N_r$ which can be identified as the
`Hall conductance' and which provides information on the
renormalization of the theory at large distances.\cite{PruiskenWang}
Apparently we have $m'\!=\!m$. The same conclusion can be drawn for
the $\qo$ parameter (i.e.
$\qo'\!=\! \qo\tr \qL\langle Q\rangle/2N_r\!=\!\qo$) and hence we are
dealing with a critical fixed point theory!
The full significance of this result will become clear in the
forthcoming sections where we make contact with the theory of chiral
edge bosons.

For the remainder of this section we will elaborate on several other
identities and relations that will be used later on. The most
important pair correlation of the $Q$-fields can be obtained as
follows,
\bea
	N(x,x')&=& \pi^2\qr_{\rm edge}^2\langle Q_{+-}^{\qa\qb}(x)
	Q^{\qb\qa}_{-+}(x')\rangle
	=\sum_{jj'}G_-^{jj'}(x,x')G_+^{j'j}(x',x) \nn\\
	&=& mG_-(x,x')G_+(x',x).
\label{NisGG}
\eea
Here, the $\qa,\qb$ are fixed but arbitrary replica channels and
\be
	G_-(x,x')G_+(x',x)=\frac{i}{\vd}\int\!\frac{dk}{2\pi}\cdot
	\frac{e^{ik(x'-x)}}{\vd k +2i\qo}
	=\frac{1}{\vd^2}\qy(x'-x)\exp \left[-\frac{2\qo}{\vd}(x'-x)
	\right].
\label{GminGplus}
\ee
The step function $\qy$ shows that a chiral electron, being created at
position $x$ and drifting in the positive direction, can only be
destroyed at a `later' position $x'\! >\! x$. Notice that we have the
standard sum rule
\be
	\intd{x'}N(x,x')=\pi\qr_{\rm edge}/\qo
\ee
The other pair correlations of the $Q$-fields vanish
identically. In particular, it is straightforward to show that
\be
	\left\langle Q^{\qa\qb}_{pp}(x)Q^{\qg\qd}_{p'p'}(x')
	\right\rangle_{\rm cum}=0
\label{QQcumnul}
\ee
for all $p,p'\!=\!\pm$ and all replica channels $\qa,\qb,\qg,\qd$.
Next we wish to clarify the significance of several $Q$-field
operators that have appeared in different contexts before.
First, there are the higher order corrections to the theory of
(\ref{Ssplit}) of the type (see Appendix~C)
\be
	\tr[\fr{m}{2}\prt_x+\pi\qo\qr_{\rm edge}\qL,Q]^2.
\label{type1}
\ee
Secondly, we mention the bilinear combinations of the form
\be
	A_1 \tr\qL Q\; \tr\qL Q+A_2 \tr[\qL,Q][\qL,Q]
\label{bilinears}
\ee
which are known to describe the anomalous fluctuations in the density
at the quantum Hall transitions, as well as in the localization
problem in $2\!+\!\qe$ dimensions.\cite{PrPRB85}
We have already seen, however, that the density of chiral electrons
does not fluctuate as one moves along the edge
and we therefore expect (\ref{bilinears}) to be
irrelevant. A classification of these operators follows from the
classical equations of motion of the topological action
(\ref{Ssplit}) which can be written as
\be
	[\fr{m}{2}\prt_x+\pi\qo\qr_{\rm edge}\qL,Q]=0.
\label{mottop}
\ee
This immediately implies that the higher dimensional operators
(\ref{type1}) are, in fact, {\em redundant}. Next, from the identity 
\be
	\int^x\! dx'\; \tr\qL Q(x')\; 
	\tr[\fr{m}{2}\prt_{x'}+\pi\qo\qr_{\rm edge}\qL,Q(x')]\qL=0
\ee
it directly follows that the first term in (\ref{bilinears}) is
redundant as well. Finally, from (\ref{mottop}) one also obtains
\be
	\int^x\! dx' \;
	\tr[\qL,Q(x')]\left[\fr{m}{2}\prt_{x'}+\pi\qo\qr_{\rm edge}\qL,
	[\qL,Q(x')] \right]=0,
\label{lasttrace}
\ee
and it is readily seen that the second operator in
(\ref{bilinears}) is also redundant.

\subsection{Plateau transitions revisited}
\label{secplatrev}
\subsubsection{Long range potential fluctuations}
In this section we show how the notion of critical edge states can be
used in order to gain insight into the problem of `long
ranged potential fluctuations'. This longstanding problem, which is
very difficult to handle within the formal non-linear sigma model
methodology, plays an extremely important role experimentally. For
instance, it has been stressed many times and at many places elsewhere
that the plateau transitions as observed in the detailed experiments
of H.P. Wei {\it et al.} \cite{HPWei} 
are very difficult to observe in general in
arbitrary samples, due to the presence of slowly varying potential
fluctuations. 

A slowly varying potential is the generic type of disorder in the
standard GaAs heterostructure, which has historically led to
semiclassical considerations (percolation picture) of delocalization
near the Landau band center.\cite{PrangeinPrange} 
It is important to recognize that also our
critical system (\ref{condscaling}) is very sensitive to the presence of smooth
potentials (or ``inhomogeneities'') in the sample. For example, the
critical magnetic field $B^*$ may be slowly varying throughout the
system due to inhomogeneities in the electron density. This means that
the scaling result is valid only up to a certain fixed value for
$L$. Beyond this value the remaining ``extended'' states in the
problem may be confined to the equipotential contours of the
inhomogeneity potential, quite
similar to the semiclassical picture of percolation.

It is generally difficult to obtain detailed knowledge on the various
length- and energy scales that are involved in the cross-over problem
between percolation and localization.
In what follows, we present the simplest possible scenario for
crossover that enables us to deal simultaneously with interaction
effects and such basic concepts
as `mean field theory' and `universality' of the plateau transition.

\subsubsection{Quantum percolation}
\begin{figure}
\begin{center}
\setlength{\unitlength}{1mm}
\begin{picture}(140,65)(0,0)
\put(0,5)
{\epsfxsize=130mm{\epsffile{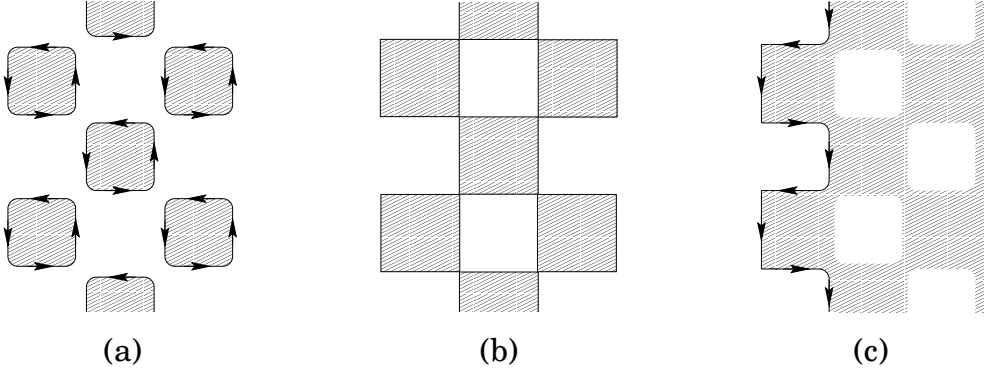}}}
\end{picture}
\caption{
Backbone cluster as a network of saddlepoints. 
Shaded areas have 
$\nu\! =\! 1$, white areas $\nu\! =\! 0$. The arrows indicate the
direction of the currents.
(a) Less than half filling; (b) exactly half filling;
(c) filling fraction larger than one half.}
\label{figpercolat}
\end{center}
\end{figure}

In order to fix the thought, we imagine the equipotential contours
near half filling to form a large cluster (Fig.~\ref{figpercolat}).
\cite{Chalker}

\begin{figure}
\begin{center}
\setlength{\unitlength}{1mm}
\begin{picture}(80,80)(0,0)
\put(0,0)
{\epsfxsize=80mm{\epsffile{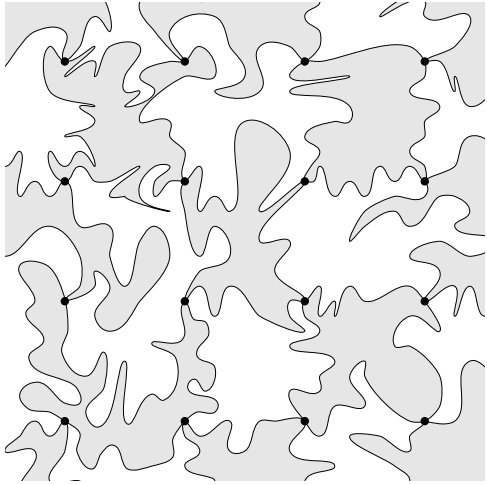}}}
\end{picture}
\caption{Backbone cluster as in fig.~\ref{figpercolat}b, but with highly
ramified contours between saddle points~($\bullet$).}
\label{figramified}
\end{center}
\end{figure}

Since the disconnected, closed contours do not contribute to the
transport, we focus our attention to an infinite backbone cluster
which we take as a regular 2D array of `saddlepoints' and we
disregard all the loose hanging, finite pieces (Fig.~\ref{figpercolat}b).
The saddlepoints (the sites of the square lattice) are connected to one
another by the disordered 1D chiral edge channels (links on the lattice).
This network can alternatively be looked upon as a checkerboard with
filling fractions alternating between the values 
$\nu\! =\! 0$ and $\nu\! =\! 1$.
The kinetic part of the action for this system may be written in the
form of (\ref{Sgap})
\be
	S[Q]=\fr{1}{8}\intdxx m(\vec x)\tr \qe_{ij}Q\prt_i Q\prt_j Q
\label{Snetkin}
\ee
with $m(\vec x) \! =\! 0,1$ (Fig.~\ref{figpercolat}b).
Using the parametrization of (\ref{Tut}) the action can also be
written in the form (\ref{Ssplit}) which is now solely defined on the
links of the square lattice,
\be
	S[Q]=2\pi i\cdot q[U]+\half\sum_i\oint_i\! dx\;
	\tr(\qL t\prt_x t\inv)+\pi\qo\qr_{\rm link}\sum_i\oint_i\! dx\;
	\tr\qL Q.
\label{linkaction}
\ee
Here, the sum is over all the black squares and the integrals are over
the contours of the black squares.
Despite the fact that this action does not contain any dissipative
($\qs_{xx}$) terms, it is easy enough to show that in the long
wavelength limit, (\ref{Snetkin}) reduces to the form of the sigma
model action (\ref{Ssigma}) with
\bea
	\qs_{xx}^0=1/2 \hskip1cm &;& \hskip1cm \qs_{xy}^0=1/2
\label{sigmahalf}
\eea
The reason for this is contained in the fact
that the saddlepoints act like scattering centers which render the
system dissipative at large distances. In order to demonstrate this,
all one needs to do is to follow up on (\ref{background})
where the background field $t_0$ now represents the `slow modes' that
are kept. The $t$ field variables are the `fast modes' which contain
all the wavelengths smaller than the lattice constant, i.e. the
average distance between the saddlepoints, and which are integrated
out.
This leads to an effective action for each link according to
\bea
	S_{\rm link}[t_0]&=& \half\int_{\rm link}\!\!\!\! dx\;
	\tr(\langle Q\rangle t_0\prt_x t_0\inv)
	+\fr{1}{8}\left\langle\left[ \int_{\rm link}\!\!\!\! dx\;
	\tr(Q t_0\prt_x t_0\inv)\right]^2
	\right\rangle_{\rm cum} \nn\\
	& = &
	\fr{1}{2}\int_{\rm link}\!\!\!\! dx\;
	\tr(\qL t_0\prt_x t_0\inv)
	-\fr{\qs_{xx}^0}{8}\int_{\rm link}\!\!\!\! dx\;
	\tr (\prt_x Q_0)^2
\label{Slink}
\eea
where $Q_0=t_0\inv\qL t_0$
and the expectation is with respect to the theory (\ref{Ssplit})
with $m\!=\!1$.
The subscript `cum' indicates that only connected diagrams are taken.
The $\qs_{xx}^0\! =\! L_0/2$ 
is the 1D conductivity of a single channel of length
$L_0$, a well-known result in the theory of pure metals.
These results are obtained by making use of (\ref{QisqL}) as well as
(\ref{NisGG}--\ref{QQcumnul}) in the limit $\qo\!=\!0$.
Next, by taking the sum over all links one can absorb the factor $L_0$ into
the definition of a 2D integral,
\be
	-\fr{1}{16}\sum_{\rm links}L_0\int_{\rm link}\!\!\!\! dx\; \tr
	(\prt_x Q_0)^2 \longrightarrow 
	-\fr{1}{16}\intdxx \tr (\nabla Q_0)^2.
\ee
Here we only used the fact that the $Q_0$ field variable varies slowly over a
distance $L_0$.
The first term in (\ref{Slink}) can be handled in a similar way. For
instance, it can be rewritten in the form of (\ref{Snetkin}) with $Q$
replaced by $Q_0$, which is then followed by taking the continuum limit
according to 
\bea
	\half \sum_{\rm links}\int_{\rm link}\!\!\! dx\; \tr
	(\qL t_0\prt_x t_0\inv) & \longrightarrow &
	\fr{1}{8}\intdxx m(\vec x)\qe_{ij}\tr Q_0\prt_i Q_0\prt_j Q_0
	\nn\\ & \longrightarrow &
	\fr{1}{16}\intdxx\qe_{ij}\tr Q_0\prt_i Q_0\prt_j Q_0
\label{xycontlim}
\eea
The result of (\ref{Slink}--\ref{xycontlim}) is identical to the
statement made in (\ref{sigmahalf}).
Notice that (\ref{sigmahalf}) 
is precisely the point where we expect the $\sigma$
model action (\ref{Ssigma}) in the limit $N_r\!=\!0$ to have a
critical phase. Hence, we 
have established a direct connection between critical 1D edge states on the
one hand and the 2D delocalization transition of the band center on the
other. It is important to stress that this connection has the following
ingredients:
\begin{enumerate}
\item
The infinite percolation cluster at the band center contains a finite density
of saddlepoints. This translates into a finite density of scattering
centers which, in turn, is responsible for making the sample diffusive
(dissipative) at large distances.
\item
The parameters $\qs_{xx}^0, \qs_{xy}^0$ (\ref{sigmahalf}) constitute a mean
field theory of the conductances which is valid for length scales $L_0$.  
This holds for any value of $N_r$ and not just for $N_r\!=\!0$.

\end{enumerate}
Without going into further detail we mention the fact that the analysis can
easily be generalized to more complicated situations. For example, the
links between the saddlepoints need not be straight lines. They can be
taken as arbitrarily complex, non-intersecting paths reflecting the
highly ramified percolation contours (Fig.~\ref{figramified}). 
The same result
(\ref{sigmahalf}) 
applies to all cases, indicating that the general result
$\sigma_{xx}^0 =1/2$ actually stands for the quantized conductance in
one dimension.

\subsubsection{Mean field theory}

\begin{figure}
\begin{center}
\setlength{\unitlength}{1mm}
\begin{picture}(120,100)(0,0)
\put(0,5)
{\epsfxsize=120mm{\epsffile{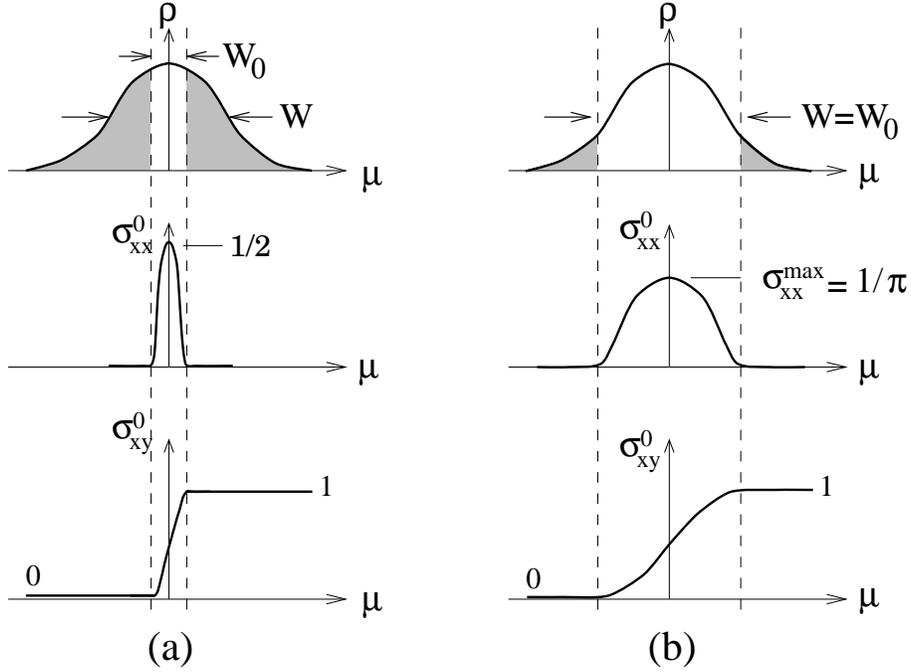}}}
\end{picture}
\caption{Mean field theory for the lowest Landau level, with varying
chemical potential $\mu$. 
(a)~Smooth long-range disorder. (b) Short-range disorder (see text).}
\label{figdos}
\end{center}
\end{figure}

Next, we wish to extend our mean field analysis (\ref{sigmahalf}) to include
also the energies away from the Landau band center. 
For this purpose we have to relate the range in energy $W_0$ within
which the equipotential contours form an infinite saddlepoint cluster
to the total bandwidth $W$ of the Landau band. It is understood that
the phrase ``saddlepoint'' actually stands for those special points
where two equipotential
contours approach each other at a distance of the order of the
magnetic length $\ell_0$ or smaller.
By assuming a simple quadratic form for the potential near saddle points 
we obtain the following estimate,
\be
	W_0\approx (\ell_0 / \ql)^2 W
\ee
where $\ql$ is the characteristic correlation length of the random
potential, which we have taken to be much larger than $\ell_0$,
and $W$ equals the amplitude of the potential fluctuations.
The sigma model theory or, equivalently, the scaling theory of
localization only applies to the (narrow) energy band $W_0$ about the
band center. For energies just outside $W_0$ the network of
saddlepoints is broken up into disconnected islands of size 
$L_0\!\times\! L_0$ (Fig.~\ref{figpercolat}a and c).
The absence of any quantum tunneling means that no correlation exists
between the islands (they are represented by independent actions as
long as one works within the free electron approach).
In the language of the $\qs$ model, the situation is represented by
putting $\qs_{xx}\!=\! 0$ but $\qs_{xy}\! =\! m\!=$integer.
The latter follows from the long-ranged correlations which still exist
near the edge and which can generally be expressed in terms of an
integer number $m$ of edge channels.
In Fig.~\ref{figdos}a we illustrate the behavior of the density of
states $\qr$ and the conductances $\qs_{ij}^0$ as a function of energy
$\mu$ at zero temperature.

The sigma model conductance parameters $\qs_{ij}^0$ can be expressed
as a function of the dimensionless quantity $\triangle\mu/W_0$
\be
	\qs_{ij}^0=f_{ij}(\triangle\mu/W_0)
\label{sigij=fij}
\ee
where $\triangle\mu$ is the energy relative to the Landau band center.
The $f_{ij}$ are non-universal and generally depend on the microscopic
details of the randomness. For comparison we have plotted the results
of the more familiar theory of short-ranged scatterers
(self-consistent Born approximation) in Fig.~\ref{figdos}b. In this
case, there is only a small difference between $W_0$ and $W$ due to 
the localized states in the Gaussian tails of the Landau band. 

An estimate for $L_0$ can be obtained as follows. 
Let $|\triangle\mu|\!\approx\! W_0$ denote the energies where the
saddlepoint breaks up into disconnected equipotential contours of size
$L_0\!\times\! L_0$ (Fig.~\ref{figpercolat}a,c).
According to the semiclassical picture of percolation we can relate
the typical cluster size $\xi_p$ to the energy $\triangle\mu$
according to
\be
	\xi_p\sim \ql(\triangle\mu/W)^{-4/3}
\label{exp4/3}
\ee
where the critical index $4/3$ is the exponent for semiclassical
localization. 
By identifying the points $|\triangle\mu|\!=\!W_0$
and $\xi_p\!=\! L_0$ in (\ref{exp4/3}) we obtain the estimate
\be
	L_0\approx\ell_0(\ql/\ell_0)^{11/3}\hskip6mm (\ql\gg\ell_0)
\label{estimateL0}
\ee
or, more generally,
\bea
	W_n\approx{{l_n}\over\lambda} W & \hskip5mm ; \hskip5mm &
	L_n\approx\ell_n(\ql/\ell_n)^{11/3}\hskip6mm (\ql\gg\ell_n).
\eea
The $\ql$ is an adjustable parameter in
the theory and it ranges between microscopic distances
($\ell_0\!\approx\! 100$\AA~) and infinity.

\subsubsection{Interaction effects}
\label{secintereff}
It is quite possible that $L_0$ (\ref{estimateL0}) 
is many times larger than the
micron regime which is the typical scale for inelastic processes at
low temperatures. This means that the critical behavior 
(\ref{condscaling}) cannot
be observed within the limitations of ordinary laboratory experiments.
This, then, is the easiest and crudest explanation for the lack of scaling
in many samples.
As a first step toward a more quantitative understanding of transport at
finite $T$, we come back to the distinction, made in the beginning,
between the backbone cluster and the disconnected, `loose
hanging' pieces.
Due to the electron-electron interactions, motion of the conducting
electrons on the saddlepoint network is affected by the localized
electrons. This may be expressed in terms of a {\it relaxation time} 
$\qt_{\rm in}$ which is a characteristic time for equilibration
between the conducting and localized electrons.
Later on in this paper (Section~\ref{sectauin}) 
we shall address the problem of
interaction effects and show that
\be
	1/\qt_{\rm in}=\qb_1 T+\qb_2 T^2 +\cdots
\label{inelastic}
\ee
at low temperatures. This expression is determined by the
collection of `nearly saddlepoints' where quantum tunneling is not possible
but where the interactions
between the conducting and localized particles are strongest nevertheless.
The importance of `nearly saddlepoints' can be seen by comparing the  
wavefunctions at different energies close to the Landau band
center.
What is a saddlepoint configuration at one energy may
turn into a `nearly saddlepoint' at another and vice versa.
These abrupt changes in the configuration of the conducting network
at slightly different energies blur the distinction between
saddlepoints and `nearly saddlepoint' configurations as far as finite
temperatures are concerned. This means that
the relaxation time $\qt_{\rm in}$ (\ref{inelastic}) 
determines an effective bandwidth
$W_{\rm eff}\! =\! W_0 +\qt_{\rm in}\inv$ 
of states that contribute to the
conduction at finite temperatures.
Eq. (\ref{sigij=fij}) 
is replaced by the expression
\be
	\qs_{ij}^0(T)=f_{ij}(\triangle\mu / W_{\rm eff})=
	f_{ij}(\triangle\mu /[W_0+\qt_{\rm in}\inv]).
\label{replacesemicl}
\ee
This result is a characteristic feature of long-ranged potential
fluctuations and it does not occur in the problem of short-ranged
scatterers. 
To conclude this section, we shall next estimate the range of validity of
the result (\ref{replacesemicl}). Write 
\bea
	v_d \tau_{\rm in} = L_{\rm in} & \hskip5mm , \hskip5mm&
	v_d \approx 2 \pi l_0^2 W/\lambda.
\eea
The $L_{\rm in}$ is the mean free path for drifting along the links
of the lattice. We mentioned earlier already that the actual path
between two saddle points is arbitrarily convoluted and very long.
Let $L_t$ denote the actual path length between saddle points, then the
criterion for scaling is clearly given by
\be
	L_{\rm in} > L_t .
\label{scalingcrit}
\ee
Next we use the ramification hypothesis\cite{Prange} in order to relate
$L_t$ to the  
shortest distance between saddlepoints ($L_0$). We obtain
\be
	L_t \propto L_0^\qs
\ee
with $\qs$ somewhere between 1 and 2.
The criterion for scaling (\ref{scalingcrit}) now implies
\be
	\tau_{\rm in}^{-1} < (l_0 / \lambda)^{8\qs/3} W_0 \ll W_0 .
\ee
This result indicates that (\ref{replacesemicl}) 
is very likely to be observed in the
(many) samples that are characterized by a smooth disorder potential. 
The results of this section are consistent with the recently reported 
empirical fitting \cite{semiclass}
of the transport data in the quantum Hall
regime. Since we are necessarily operating with an almost complete
lack of knowledge on the microscopic details of sample disorder, it is
conceivable that other types of inhomogeneity, especially those in low mobility
samples, explain the same thing.

\subsubsection{Modified $\qs$ model representation}

The subjects of critical edge states as well as long-ranged disorder have
left several conceptual questions that still need to be
answered. For example, we have seen that short-ranged disorder causes
interchannel scattering between the chiral edge states. Since we do not
expect interchannel scattering to occur when the potential fluctuations
are smooth (relative to the magnetic length), it is necessary to
re-investigate the meaning of instanton vacuum theory (\ref{Ssigma})
for $\nu\! >\! 1$ ($\qs_{xy}^0 \! > \! 1$).
Scattering between multiple edge states is avoided by writing, instead of
(\ref{Ssigma})
\bea
	S_{\rm eff}[Q^{(n)}] &=& \sum_{n=0}^\infty\left[
	-\fr{1}{8}\qs^{(n)}_{xx}
	\intdxx \tr[\nabla Q^{(n)}]^2
	+\fr{1}{8}\qs^{(n)}_{xy}\intdxx 
	\tr\qe_{ij}Q^{(n)}\prt_i Q^{(n)}\prt_j Q^{(n)} \right. \nn\\
	&& \left. +\pi\qr^{(n)}\qo \intdxx \tr \qL Q^{(n)} \right]
\label{multiQ}
\eea
where the sum runs over all the Landau levels $n$. The $Q^{(n)}$
stands for an independent field variable $Q$ for each Landau level
separately. The $\qs_{ij}^{(n)}$ are the $n$'th Landau level contributions
to the mean field conductances, which are now given by
\be
	\qs_{ij}=\sum_{n=0}^\infty \qs_{ij}^{(n)}.
\ee
The $\qs_{ij}^{(n)}$ are all the same (Fig.~\ref{figdos}a) except for an
appropriate shift in energy.
Since $0\! \leq \! \qs^{(n)}_{xy} \! \leq \! 1$ for each $n$,
it is clear that (\ref{multiQ}) is the appropriate generalization of the
theory (section~\ref{secsigmamodel}) 
to include filling fractions larger than one.
The theories of (\ref{multiQ}) and (\ref{Ssigma}) are identical as far as
the critical behavior of the plateau transitions is concerned.
Equation (\ref{multiQ}) cannot, however, be used in the limit of small
magnetic field, where the Landau levels partly or completely overlap.
The details of crossover require a separate analysis.

\subsubsection{Topological principle}

In Refs.~11, 23 a {\em topological principle} for Hall quantization
was introduced. The basic idea is to relate 
the concept of dynamic mass generation in
asymptotically free field
theories to the quantization of the Hall conductance, which is now
recognized as a universal quantum phenomenon at macroscopic length
scales. The formulation presented in Refs.~11, 23 is actually
incomplete because the subtleties edge effects were not sufficiently
understood 
at that time. In order to see whether the instanton vacuum approach
is, in fact, free of ambiguities, we shall follow up on the background
field method which is known to generate the Kubo formulae for the
conductances. Write
\be
	\exp S_{\rm eff}[t_0] =
	\int\!{\cal D}[Q]\exp \left(S_0[t_0\inv Qt_0]
	+\pi\qr_0\qo\Tr\qL Q
	\right)
\label{insertion}
\ee
where 
\be
	S_0[Q]=-\fr{1}{8}\qs_{xx}^0\Tr(\nabla Q)^2+\fr{1}{8}\qs_{xy}^0
	\Tr \qe_{ij}Q\prt_i Q\prt_j Q.
\label{insertion2}
\ee
Eq. (\ref{insertion}) defines an effective action $S_{\rm eff}$ for the
fixed and slowly varying matrix field $t_0$. One can show that $S_{\rm
eff}$ is of the same form as $S_0$, i.e.
\be
	S_{\rm eff}[t_0]=-\fr{1}{8}\qs_{xx}\Tr(\nabla Q_0)^2
	+\fr{1}{8}\qs_{xy}\Tr \qe_{ij}Q_0\prt_i Q_0\prt_j Q_0
\label{SeffS0}
\ee
with $Q_0\!=\!t_0\inv \qL t_0$.
Eq.~(\ref{SeffS0}) is actually the only possible action that respects
the global $U(2N_r)$ symmetry  as well as the local 
$U(N_r)\!\times\! U(N_r)$ gauge invariance of the problem. The main
problem next is to obtain explicit knowledge of the ``effective''
parameters $\qs_{ij}$ in (\ref{SeffS0}) which now represent the
(exact) Kubo expressions for the conductances. As long as one works
with spherical boundary conditions on the matrix field $Q$ (which have
been assumed from the start), the quantization of the Hall conductance
is readily established. All that one needs is in fact that the theory
develops a mass gap in the limit of large distances. The insertion of
slowly varying background fields (with $Q\!=\!\qL$ at the edge) should
then leave the theory unchanged in the limit $\qo\!\naar\!0$. This,
then, directly leads to the statement saying that
$\qs_{xx}\!=\!0$ and $\qs_{xy}\!=$integer.

The renormalization group flows, obtained from instanton calculations,
can next be used to show how the conditions of the quantum Hall effect
appear as stable, infrared fixed points of the theory for arbitrary
number of field components $N_r$.

Although spherical boundary conditions are naturally
imposed on the weak coupling problem due to the finite action
requirement of 
topological excitations, they are, however, 
controversial in the strong coupling regime.

Armed with the insight gained from edge excitations in the previous
sections, we next apply the background field procedure to the theory,
but now with free boundary conditions on $Q$, as it should be. For the
special case where the Fermi energy lies in a density of states gap,
(\ref{insertion}) has already been addressed in Section~B2.
$S_{\rm eff}$ for arbitrary $N_r$ is given by
\be
	S_{\rm eff}[t_0]=2\pi i m\cdot q[Q_0]-\fr{m^2}{32\pi\qo\qr_{\rm edge}}
	\ointdx \tr (\prt_x Q_0)^2
\label{SQ0}
\ee
where 
$\;\; q[Q_0]\! =\!  \fr{1}{16\pi i} \Tr\qe_{ij}Q_0\prt_i Q_0\prt_j Q_0$
and the contour integral is along the sample edge.
Comparing (\ref{SQ0}) with (\ref{SeffS0}) we see that the quantum Hall
conditions are satisfied, but there are additional edge terms which
are clearly the result of the chiral edge modes in the problem.
Eq.~(\ref{SQ0}), in the limit $\qo\!\naar\!0$, forces the background
field to obey the classical equations of motion (defined along the
sample edge)
\be
	\prt_x Q_0=0.
\label{classeqm}
\ee
The solution $Q_0\!=$constant at the edge simply means that spherical
boundary conditions are automatically enforced by the chiral edge
excitations. Notice that the effect of $S_{\rm eff}$ reduces to that
of a phase factor which is immaterial provided the Hall conductance
$m$ precisely equals an integer. Physically, this phase factor arises
from an integer number of edge electrons that have crossed the Fermi
level as a result of the background field insertion.\cite{PrinPrange}



The same procedure can
be repeated for the theory with $\qs_{xx}^0\!\neq\!0$, making use of
the fact that a mass gap exists in the system of long wavelength
excitations, i.e. a finite localization length $\xi$. One expects
(\ref{SQ0}) to be modified according to
\be
	S_{\rm eff}[t_0]=-\fr{\qs_{xx}}{8}\Tr(\nabla Q_0)^2
	+ 2\pi i \qs_{xy}q[Q_0]
	-    
	g_m L_\qo\ointdx \tr(\prt_x Q_0)^2
\label{St0g}
\ee
where the $\qs_{ij}$ represent the `conductances'
\bea
	\qs_{xx}=f_{xx}(\qo\xi^2)\approx {\cal O}(\qo\xi^2) 
	\hskip1cm &;& \hskip1cm
	\qs_{xy}=f_{xy}(\qo\xi^2)\approx m +{\cal O}(\qo^2\xi^4).
\label{sigetaxi}
\eea
Here, $g_m\! =\! m/2$ is the quantized 1D conductance of the chiral
edge states and 
$L_\qo\!=\!m/16\pi\qo\qr_{\rm edge}$ is the frequency induced length scale. 
In the limit $\qo\!\naar\!0$ the $Q_0$ entering (\ref{St0g}) is forced
to obey not only the classical equations of motion on the edge
(\ref{classeqm}), but also those arising from the bulk kinetic term in
(\ref{St0g}). The solutions are known as {\em instantons} and just as
has happened before in the trivial example with a density of states
gap in the bulk, $S_{\rm eff}$ is immaterial as long as 
$\qs_{xx}\!=\!0$ and $\qs_{xy}\!=$integer.
Therefore, the quantum Hall effect can be understood in terms of a
continuous symmetry which is dynamically restored in the limit of
large length scales.

\vskip4mm

In summary we can say that the ``quantum Hall effect'' is a robust and
general feature of the instanton vacuum theory for all values of
$N_r$. Our theory of topological quantum numbers is based on 
two general assumptions only, namely the existence of a mass gap in
the bulk as well as massless excitations at the edge. Both are valid
for the $\qs$-model in two dimensions for all (non-negative) values
of~$N_r$.  

The results of this section can be used to demonstrate that a
phase transition must occur when $\qs_{xy}^0$ passes through
half-integer values (or the instanton parameter $\qy$ passes through
$\pi$). The argument\cite{PrinPrange} is based on the fact that the
Hall conductance 
$\qs_{xy}$ must make an integer step when $\qs_{xy}^0$ is approached
from the integer sides.
These phase transitions separate the different instanton vacua which
are now labeled by macroscopic quantum numbers
(i.e. $\qs_{xy}\!=$integer) and they are distinct from each other by
the number of massless modes that exist near the edge of the system.
Apart from the close contact with quantum Hall physics, the argument
for a phase separation between the different instanton vacua proceeds
along similar lines as 't~Hooft's duality argument.\cite{Toeft}

Finally, we mention that the results of this paper have interesting
consequences for the idea of having a first order phase transition at
$\qy\!=\!\pi$ (as found e.g. in the large $N$ theory of the $CP^N$
model\cite{Ed}).
First order instabilities provide an alternative physical scenario of
Hall quantization and will be discussed elsewhere.\cite{corrolaries}

\ns{Derivation of the full edge theory}
\label{secderivfull}

\subsection{Preliminaries}
\label{secprelim}

From now onward we turn to the fermionic path integral. Following [I],
one can formulate a complete theory of $Q$-matrix fields that includes
external potentials as well as interactions by making use of such
concepts as `smallness', ${\cal F}$-invariance and ${\cal F}$-algebra.
We will proceed by summarizing the main ingredients of the fermionic
path integral approach (Sections~A and B). In Section~C we present the
main steps of a derivation of $Q$-field theory at the edge, assuming
that the Fermi energy lies in a Landau gap. The various manipulations
closely follow the effective action procedure for free electrons, and
we refer to the original works of Refs.~8 and 11 for the missing
details. 

\subsubsection{Notation}

Let us start by writing down the $Q$-field theory for disordered
electrons in 2+1 dimensions in the presence of Coulomb interactions
and external potentials, derived in  [I] 
\bea
	S[A,\tQ,\ql] & = &
	-\fr{1}{2g}\Tr \tQ^2 +\Tr\ln[i\qo+i\hat{A}_0
	+i\hat{\ql}
	+\mu-\hat{\cal H} +i\tQ] \nn\\
	& & 
	-\half\qb\int\! d^2 x d^2 x'
	\; \ql\dagg(x)U_0\inv(x-x') \ql(x').
\label{S[Q,lambda]}
\eea
The symbols appearing in this action have the following meaning:
The $\tQ(\vec x)$ is an infinite-dimensional  matrix field with two
replica indices and two 
Matsubara frequency indices. (In the derivation of the above  action, it
arises as a quadratic expression in the original electron field
$\psi$; The saddle point is given by
$\tQ^{\qa\qb}_{nm}\propto\psi^\qa_n\bar\psi^\qb_m$.)
Upper Greek indices denote a replica channel, running from 1 to $N_r$,
while lower Latin indices stand for Matsubara frequencies,running from
$-\infty$ to $\infty$.
The matrix field $\tQ$ can be split into `transverse' and
`longitudinal' components
\bea
	\tQ=T\inv PT \hskip0.5cm
	& P=P\dagg  \hskip0.5cm
	& T\in SU(2\tilde{N}).
\label{defQtilde}
\eea
Here, $P$ has only block-diagonal components in frequency space 
(i.e. $P^{\qa\qb}_{nm}\!\neq\! 0$ only for 
$\qo_m\qo_n\! > \! 0$) and $T$ is a
unitary rotation. The size of the $\tQ$-matrix is given by
$2\tilde{N}$, namely the number of replicas times the size of
Matsubara frequency space.
The matrix $\qo$ is unity in replica space, while in frequency space
it is a diagonal containing the fermionic frequencies,
\bea
	\qo^{\qa\qb}_{nm}=\qd^{\qa\qb}\qd_{nm}\qo_n \hskip1cm &;& \hskip1cm
	\qo_n=\fr{2\pi}{\qb}(n+\half)
\eea
with $\qb$ the inverse temperature.
The symbol `Tr' denotes a matrix trace as well as spatial integration.
All spatial integrals are taken in the upper half plane 
$y\! >\! 0$. The
sample edge is given by the line $y\! =\! 0$.
The $U_0\inv(\vec x-\vec x')$ is the matrix inverse of the Coulomb
interaction $U_0(\vec x-\vec x')$.
$A_\mu$ is the external potential; $\ql$ is the plasmon field. It is
assumed that these fields do not have a static ($n\! =\! 0$) component.
The `hat' notation ($\widehat{\hphantom H}$) appearing in
(\ref{S[Q,lambda]}) is defined as follows
\be
	\hat x=\sum_{\qa=1}^{N_r}\sum_{n=-\infty}^\infty
	x^\qa_n \tIan
\label{defhat}
\ee
where $\tIan$ is the unity matrix in the $\qa$'th replica channel,
shifted by $n$ places in frequency space
\be
	(\tIan)^{\qb\qg}_{kl}=\qd^{\qa\qb}\qd^{\qa\qg}\qd_{k-l,n}.
\ee
The ${\cal H}$ is the kinetic energy (differential) operator,
\bea
	{\cal H}=\fr{1}{2\me}(\pileft-\vec A)\cdot(\vec\pi-\vec A)
	\hskip0.7cm
	& \vec\pi= \fr{1}{i}\nabla-\vec A_{\rm st} \hskip0.7cm
	& \pileft=-\fr{1}{i}\nablaleft-\vec A_{\rm st}
\eea
where $\vec A_{\rm st}$ describes the static magnetic field according
to
$\curl\vec A_{\rm st}\! =\! B_{\rm st}$.

\subsubsection{Flux-charge composites}
In order to describe the FQHE one also needs to include a {\em
statistical} or Chern-Simons gauge field $a_\mu$
in~(\ref{S[Q,lambda]}) as follows
\be
	S[A,\tQ,\ql] \naar S[A+a,\tQ,\ql] 
	+ \fr{i}{8p\pi}\intd{\qt d^2 x}
	\qe^{\mu\nu\qs}a_\mu\prt_\nu a_\qs,
\ee
with $\qe^{\mu\nu\qk}$ the antisymmetric tensor in 2+1 dimensions and
$2p$ an even integer denoting 
the number of elementary flux quanta $h/e$ attached to every
electron. 
Note that in this procedure the zero-frequency components of all
fields
are to be treated at a mean field level.
This amounts to
adding an extra contribution $\vec a_{\rm st}$ to the static part of
the external vector potential $\vec A_{\rm st}$, resulting in an
effective magnetic field 
$B_{\rm eff}\!=\!\curl(\vec A_{\rm st}\!+\!\vec a_{\rm st})\!=\!B_{\rm st}
\!+\!2p n_{\rm e}h/e$,
with $n_{\rm e}$ the mean electron density.
Jain's composite fermion mapping is then implemented by integrating out the
field $a_\mu$. In this paper, however,  we only consider the integer quantum
Hall effect; we deal with the fractional effect in a subsequent publication.

\subsubsection{Landau gap}
A theory for the edge is obtained by
choosing the
chemical potential $\mu$ approximately halfway between Landau energies, where
the bulk density of states is virtually zero if the disorder is not
too strong.
The saddle point equation for $\tQ$ is given by
\be
	\tQ_{\rm sp}\propto \qr T\inv \qL T
\ee
where $\qr$ is the density of states and $\qL$ is the matrix appearing
in (\ref{defQ22}) but now with full frequency dependence
\be
	\qL^{\qa\qb}_{kl}=\qd^{\qa\qb}\left[
	\matrix{1 & 0\cr 0 & -1}\right]_{kl}.
\ee
Since we are interested in the limit $\qr\!\naar\!0$, we may replace
the full expression for $\tQ$ (\ref{defQtilde}) by a much simpler one,
\be
	\tQ\naar \qe T\inv \qL T, \;\;\;\;\; \qe\ll 1.
\label{newQtilde}
\ee
From detailed earlier work\cite{PrinPrange} we know that (\ref{defQtilde})
and (\ref{newQtilde}) give rise to identical results as long as the
bulk density of states $\qr$ can be safely taken to zero. However, in
order to deal with the complications of $U(1)$ gauge invariance
(section B), there is considerable advantage in working with the
simplified expression (\ref{newQtilde}), and we will refer to the
details of more elaborate analyses only when necessary.

\subsection{Gauge invariance and truncation of frequency space}
\label{secgaugeinv}

The electromagnetic $U(1)$ gauge transformations in this theory are
generated by the $\tI$-matrices. Multiplication of these matrices is
very simple,
\be
	\tIan\tI^\qb_m=\qd^{\qa\qb}\tI^\qa_{n+m},
\label{tItI}
\ee
and they form an abelian algebra. Gauge transformations are given by
\bea
	A_\mu\naar A_\mu+\prt_\mu\chi \hskip1cm &;& \hskip1cm
	\tQ\naar e^{i\hat\chi}\tQ e^{-i\hat\chi}
\label{U1transf}
\eea
with $\chi^\qa_0\! =\! 0$.
The gauge invariance of (\ref{S[Q,lambda]}) is easily checked by
writing the transformed Tr ln in the form
 $\Tr\ln(
e^{-i\hat{\chi}}[\cdots]e^{i\hat{\chi}})$, noting that
\be
	e^{-i\hat{\chi}} \qo e^{i\hat{\chi}}  =  \qo -\widehat{\prt_0\chi}
\label{rotomega}
\ee
and
\be
	e^{-i\hat{\chi}}(\vec{\pi}-\hat{\vec{A}}-\nabla\hat{\chi}) 
	e^{i\hat{\chi}}  = 
	\vec{\pi}-\hat{\vec{A}}.
\label{rotH}
\ee
In order to facilitate the expansion of the Tr ln term 
in (\ref{S[Q,lambda]}) we perform a
gauge transformation that sets $A_0\! +\!\ql\! =\! 0$.
Introducing the notation
\bea
	\widetilde{W}=\exp\left(\sum_{\qa}\sum_{n\neq 0}
	\frac{(A_0+\ql)_n^{\qa}}{\nu_n}\tI_n^{\qa}\right) \hskip1cm &;& \hskip1cm
	\vec{z}_n^{\qa}=\vec{A}_n^{\qa}
	-i\frac{\nabla(A_0+\ql)_n^{\qa}}{\nu_n}
\label{Wrot}
\eea
with $\nu_n\!=\!2\pi n/\qb$,
and the gauge invariant quantity 
$\widetilde{R}\! =\!\widetilde{W}\tQ\widetilde{W}\inv$, 
the Tr ln can be written as
\be
	\Tr\ln[i\qo+\mu -\fr{1}{2\me}(\pileft-\hat{\vec{z}})
	\cdot(\vec{\pi}-\hat{\vec{z}})+i\widetilde{R}].
\ee
Notice that 
$\vec{z}^\qa_n\! =\! i(\prt_0\vec A\! -\!\nabla A_0)^\qa_n/\nu_n$, 
from which it is clear that $\vec z$ is also gauge invariant.

\vskip0.3cm

As was the case in  [I] , we have to impose a cutoff on the size
of Matsubara frequency space. Instead of being infinite, all matrices
are now of size $2N_{\rm max}'\!\times\! 2N_{\rm max}'$ in frequency
space. The Matsubara indices sit in the interval 
$(-N_{\rm max}',\cdots,N_{\rm max}' \! -\! 1)$. 

The truncated version of the $\tI$-matrices is denoted by $\Ian$. The
`hat' notation is now defined with respect to the truncated matrices $\Ian$.
These no longer
span an abelian algebra; instead their commutators are given by
\bea
        (\Ian {\rm I}^\qb_m)^{\mu\nu}_{kl}=(\tI^\qa_n \tI^\qb_m)^{\mu\nu}_{kl}
        g_{l+m} \hskip1cm &;& \hskip1cm
        [\Ian, {\rm I}^\qb_m]^{\mu\nu}_{kl}=\qd^{\qa\qb\mu\nu}\qd_{k-l,m+n}
        (g_{l+m}-g_{l+n})
\label{I,I}
\eea 
where $\qd^{\qa\qb\mu\nu}$ means that all replica indices have to be the
same, and $g_i$ is a step function equal to one if 
$i\in\{ -N_{\rm max}',\cdots,N_{\rm max}' \! -\! 1   \}$ 
and zero otherwise.

\begin{figure}
\begin{center}
\setlength{\unitlength}{1mm}
\begin{picture}(115,35)(0,0)
\put(0,0)
{\epsfxsize=35mm{\epsffile{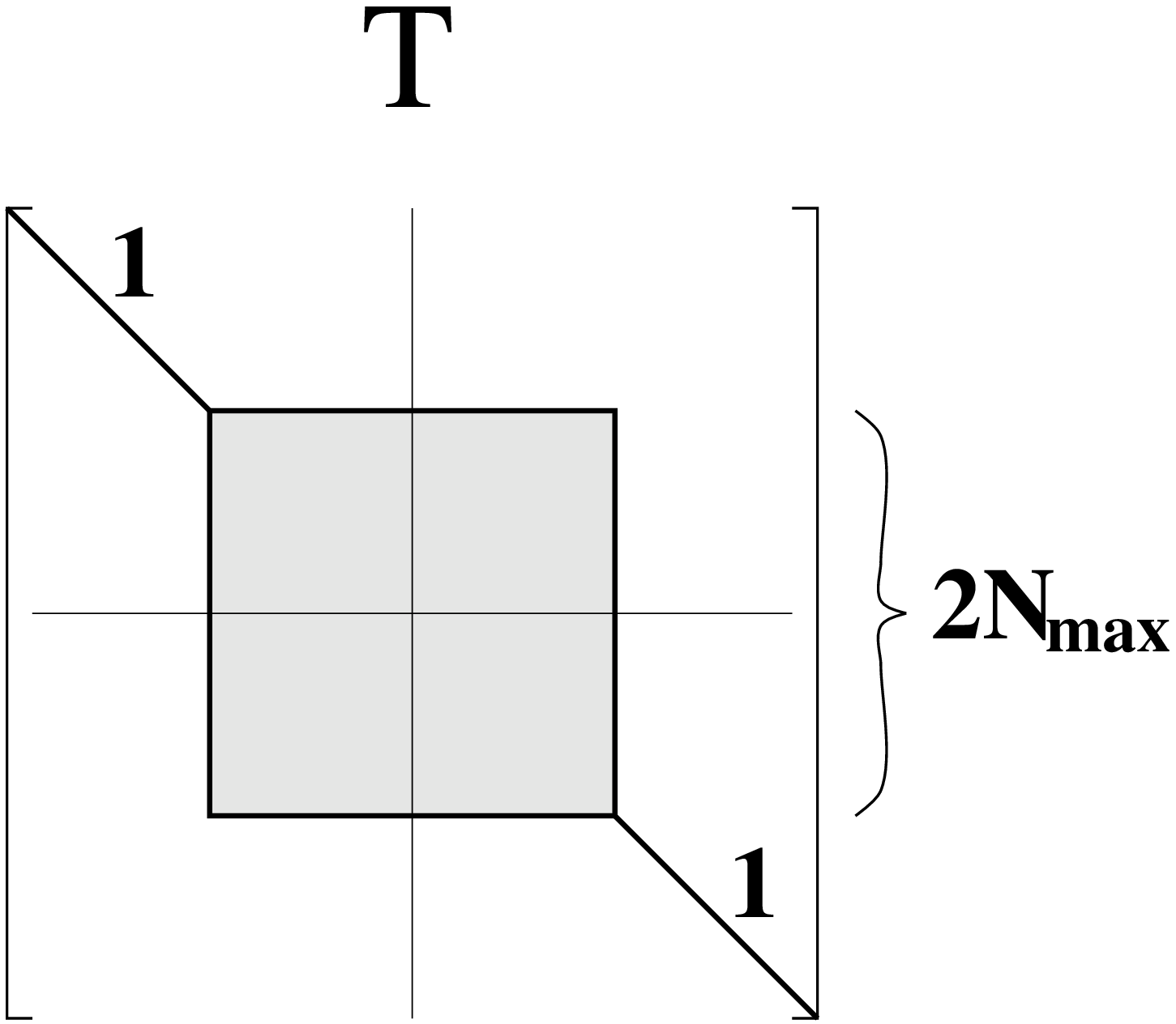}}}
\put(45,0)
{\epsfxsize=35mm{\epsffile{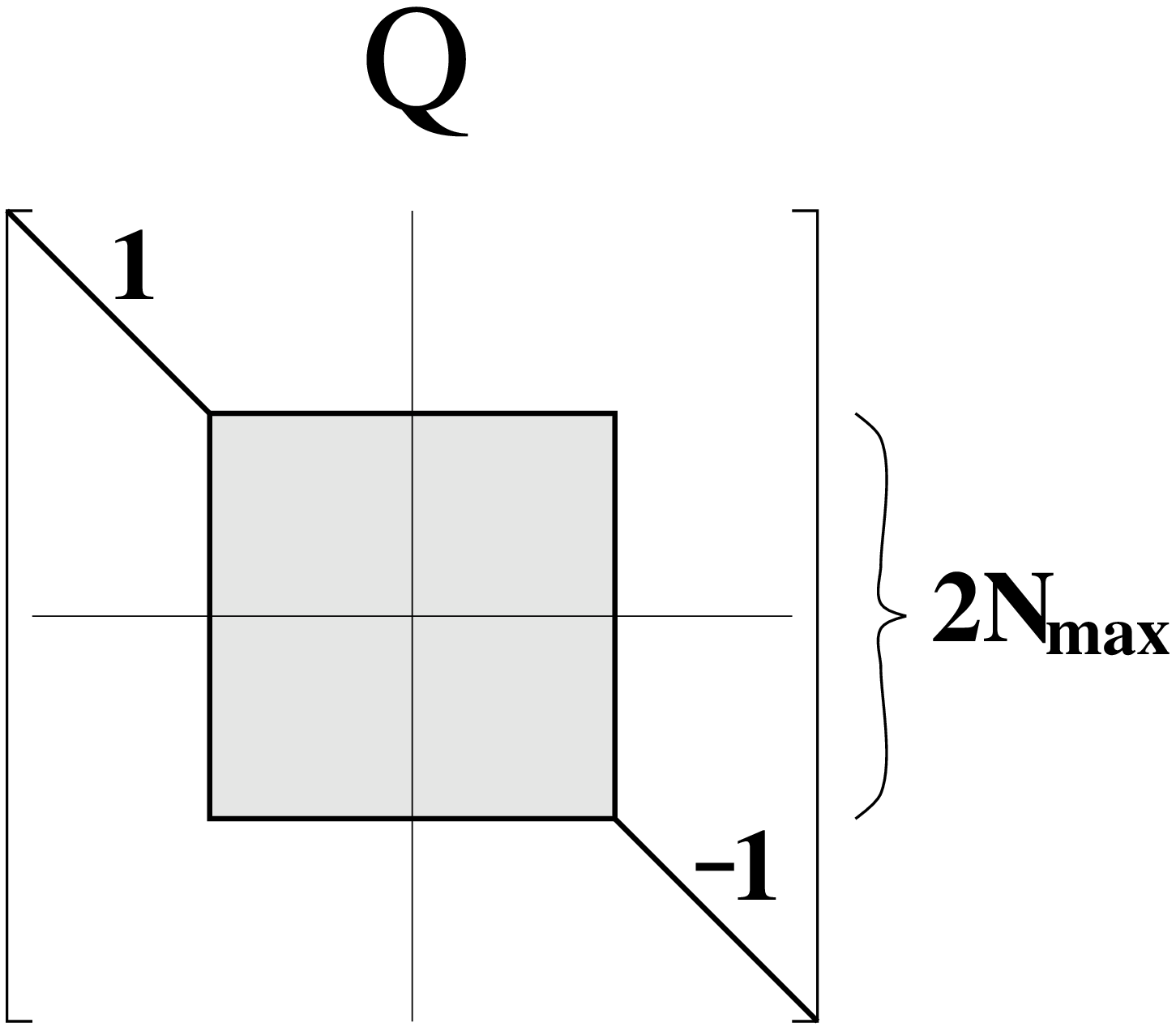}}}
\put(90,0)
{\epsfxsize=25mm{\epsffile{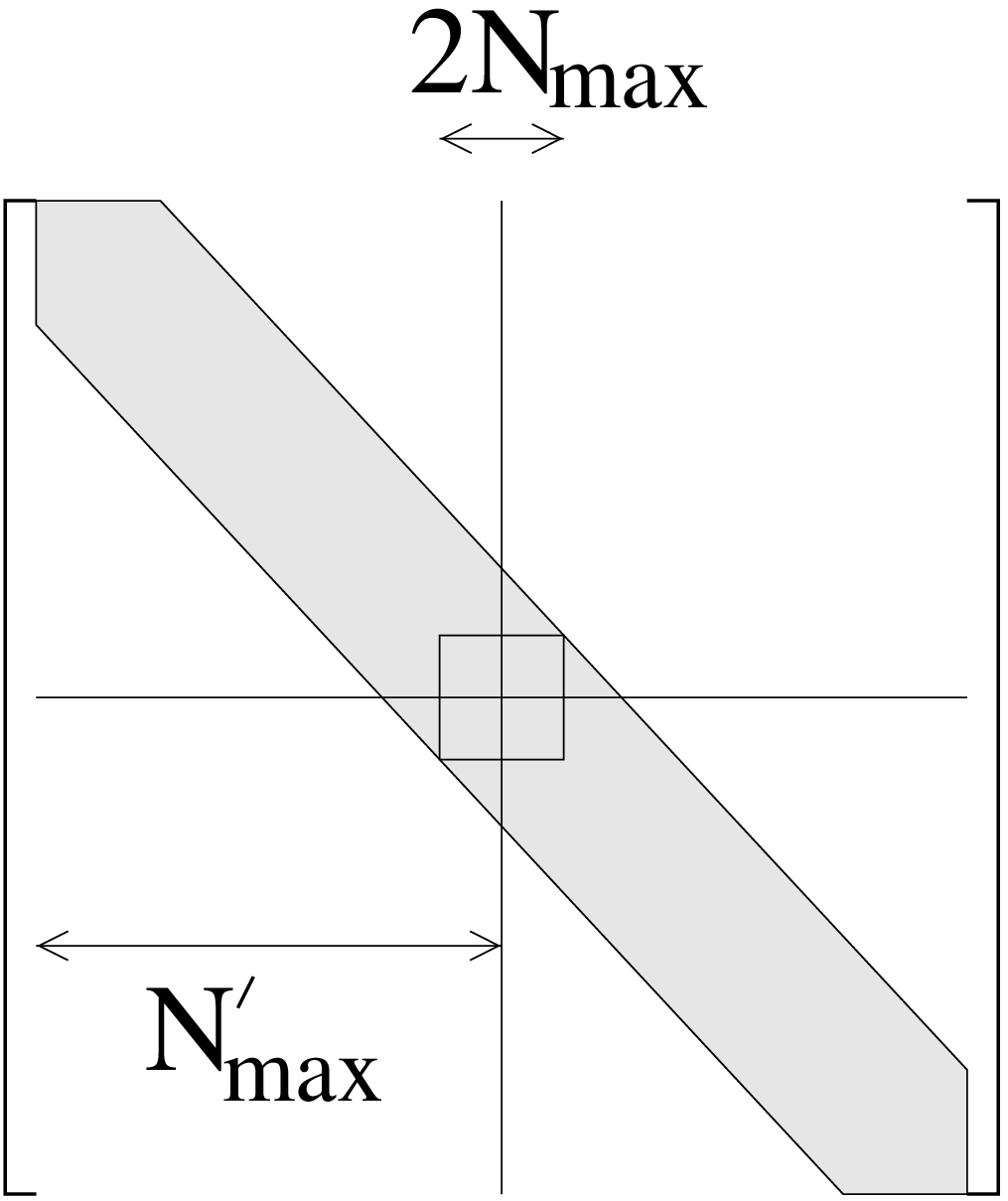}}}
\end{picture}
\caption{The truncated matrices $T$ and $Q$; 
Also is drawn the frequency band in which
$\tr\Ian Q\!\neq\! 0$}
\label{figTQIrange}
\end{center}
\end{figure}

In order to retain some form of gauge invariance, a second cutoff
$N_{\rm max}\!\ll\! N_{\rm max}'$ 
is introduced for the matrix field $T$. With the
truncated $T$ we define the truncated equivalent of $\tQ$
(see Fig.~\ref{figTQIrange}),
\be
	Q=T\inv\qL T.
\label{defQ}
\ee

It was shown in  [I]  that most of the problems caused by the
change from (\ref{tItI}) to (\ref{I,I}) can be avoided by the
introduction of the second cutoff. A remnant of the $U(1)$ symmetry is
kept in this way: invariance of the action under the truncated
equivalent of (\ref{U1transf})
\bea
	A_\mu\naar A_\mu+\prt_\mu\chi \hskip1cm &;& \hskip1cm
	Q\naar e^{i\hat\chi}Q e^{-i\hat\chi}.
\label{Ftransf}
\eea
We do not have a full symmetry
of the theory, since the integration measure ${\cal D}Q$ is {\em not}
invariant under (\ref{Ftransf}). Only in the limit of 
$N_{\rm max}\!\naar\!\infty$ is full symmetry obtained. 
It is always understood implicitly that this limit is taken in the end.


By taking the second cutoff $N_{\rm max}$ we also restrict the
interval of the frequency indices on $\ql^\qa_n$ and $(A_\mu)^\qa_n$
to $n\in [-2N_{\rm max}\! +\! 1,2N_{\rm max}\! -\! 1]$ (see
Fig.~\ref{figTQIrange}).
This interval corresponds to $\tr\Ian Q\!\neq\! 0$.

For the calculations that follow, it is convenient to work in a gauge
in which the combination $A_0\!+\!\ql$ vanishes.
For this purpose we introduce the following abbreviations
\bea
	{A}_{\mu}'={A}_{\mu}+\qd_{\mu 0}\ql \hskip0.5cm &
	W=\exp\left[\sum_{n\qa}\Ian({A}_0')^\qa_n/\nu_n\right] 
	\hskip0.5cm & 
	R=WQ W\inv,
\eea
and the action can be written as (up to a constant)
\bea
	S[Q,\ql,{A}] &=& -\half\qb\int\! d^2 x d^2 x'
	\; \ql\dagg(\vec x)U_0\inv(\vec x-\vec x')\ql(\vec x') \nn\\
	&& +\Tr\ln\left[i\qo+\mu -\fr{1}{2\me}(\pileft-\hat{\vec{z}})
	\cdot(\vec{\pi}-\hat{\vec{z}})+i\qe R\right]
\label{truncaction}
\eea
with $\vec z$ defined according to 
\be
	\vec z_n^{\;\qa}=\vec A_n^\qa-i\nabla (A_0')_n^\qa/\nu_n
\ee
with $\nu_n$ the bosonic frequency $2\pi n/\qb$.

\subsection{Expansion of the Tr ln}
\label{secexptrln}

Let us look at the last term in (\ref{truncaction}),
$X\! =\!\Tr\ln[i\qo\!+\!\mu\!-\!{\cal H}_z\!+\!i\qe R]$.
Introducing the notation
\bea
	D_{\qo}=TW\inv\qo WT^{-1} \hskip1cm &;& \hskip1cm 
	D_{\vec{z}}=TW\inv(\fr{1}{i}\nabla-\hat{\vec{z}})WT^{-1}
\eea
(where $D_{\vec{z}}$ is not a differential operator)
we can write
\be
	X=\Tr\ln[iD_{\qo}+\mu+i\qe\qL-\fr{1}{2\me}
	(\pileft\cdot\vec{\pi}+\pileft\cdot D_{\vec{z}}
	+D_{\vec{z}}\cdot\vec{\pi}+D_{\vec{z}}^2)].
\label{A}
\ee
Expansion to first order in $D_{\qo}$ and $D_{\vec{z}}$ yields
\be
	X\approx \Tr\ln G_0^{-1}+i\Tr G_0 D_{\qo}
	-\fr{1}{2\me}\Tr[G_0\pileft\cdot D_{\vec{z}}
	+G_0 D_{\vec{z}}\cdot\vec{\pi}]
\label{firstorder}
\ee
where $G_0$ is the bare Green's function 
$[\mu\!-\!\fr{1}{2\me}\pileft\cdot\vec{\pi}\!+\!i\qe\qL]^{-1}$.
The Green's function can be expressed in terms of the eigenfunctions
$\qf_{nj}$ of the bare Hamiltonian
${\cal H}_0\!=\!\fr{1}{2\me}\pileft\cdot\vec{\pi}$,
\be
	\langle x | G_0 | x\rangle =\sum_{nj}\frac{|\qf_{nj}(x)|^2}
	{\mu-E_{nj}+i\qe\qL}
\ee
\be
	\langle x |\frac{G_0\pileft+\vec{\pi}G_0}{2\me}| x\rangle =
	\sum_{nj}\frac{\qf_{nj}^{*}\fr{1}{i}\nabla\qf_{nj}
	-\qf_{nj}\fr{1}{i}\nabla\qf_{nj}^{*}-2\qf_{nj}^{*}\qf_{nj}\vec{A}}
	{2\me(\mu-E_{nj}+i\qe\qL)}.
\ee
Using the general relation 
$\qr(x)\! =\! -\fr{1}{\pi}{\rm Im}\; G^{+}(x,x)$ 
for the
density of states at the Fermi energy $\mu$, we get
\bea
	\langle x | G_0 | x\rangle =-i\pi\rho(x)\qL+c(x){\bf 1}
	\hskip5mm &;& \hskip5mm
	\langle x |\frac{G_0\pileft+\vec{\pi}G_0}{2\me}| x\rangle =
	-i\pi\vec{\jmath}(x)\qL+\vec{c}(x){\bf 1},
\eea
where $\vec{\jmath}(x)$ is the current density per energy interval
at the Fermi energy. The $c$ and $\vec{c}$ are real
functions that disappear from the last two traces in (\ref{firstorder}).
We can now write $X$ in the form
\bea
	X & \approx & \Tr\ln G_0^{-1}+\pi\intdxx \qr(x)\tr\qL D_{\qo}
	+i\pi\intdxx \vec{\jmath}(x)\cdot\tr\qL D_{\vec{z}} \nn\\
	& = & 
	\Tr\ln G_0^{-1}+\pi\intdxx \qr(x)\tr\qo R
	+i\pi\intdxx \vec{\jmath}(x)\cdot\tr[\fr{1}{i}\qL W\inv
	T\nabla (T\inv W)
	-\hat{\vec{z}}R].
\eea
Since $\mu$ lies in a gap, the density of states and the current density
are nonzero only at the edge. This means that the surface integral
becomes a line integral. If we assume constant
$\qr$ and $\vec{\jmath}$ on the edge, the resulting expression for $X$ is
\be
	X\approx \Tr\ln G_0^{-1}+\pi\qr_{\rm edge}\oint\! dx\;
	\tr\qo R-i\fr{m}{2}\ointdx
	\tr\hat{z}_x R+mS_{\rm top}[R]
\label{result1}
\ee
where we have used that 
$\fr{\prt I_{\rm edge}}{\prt\mu} \! =\!\fr{m}{2\pi}$
with the plateau-center filling fraction 
$m\!=\!\fr{n_{\rm e}}{B}\fr{h}{e}$ integer-valued,
and
$S_{\rm top}$ is the topological action 
\be
	S_{\rm top}[R]=\fr{1}{8}\Tr\qe^{ij}
	R\prt_i R\prt_j R.
\ee
Eq. (\ref{result1}), however, is not yet the complete answer. This can
be seen from a different expansion procedure which can be followed in
the special case where $T\!=\!{\bf 1}$ and $W\!=\!{\bf 1}$. 
In this case we have, instead of
(\ref{A}),
\bea
	X_2 & = & \Tr\ln[i\qo+\mu-{\cal H}_z+i\qe\qL]
	\approx \Tr\ln G^{-1}-\half 
	\Tr\left[\frac{G(\pileft\cdot\hat{\vec{z}}
	+\hat{\vec{z}}\cdot\vec{\pi})}{2\me}\right]^2 \nn\\
	&& -\fr{1}{2m_e}\Tr {\hat{\vec z}}^2 G \\
	G_{nm} & = & G_n\qd_{nm}=\qd_{nm}
	[i\qo+\mu-\fr{1}{2\me}\pileft\cdot\vec{\pi}]^{-1}. \nn 
\eea
This expression can be written as
\bea
	X_2 &\approx& \Tr\ln G^{-1}-\half \sum_{ij}\sum_{n\qa}
	\intd{^2 x d^2 x'} (z_i)_n^{\qa}(x)(z_j)_{-n}^{\qa}(x')
	(\Pi_{ij})_n^{\qa}(x,x') \nn\\
	&& - \fr{1}{2m_e}\sum_i\sum_{n\qa}\intdxx (z_i)_n^{\qa}
	(z_i)_{-n}^{\qa} \tr G(x,x).
\label{X2approx}
\eea
The `polarization operator' $\Pi_{ij}$ is given by
\bea
	(\Pi_{ij})_n^{\qa}(x,x') & = & (\fr{1}{2\me})^2
	\tr[G(x,x')(\pileft_i+\vec{\pi}_i)I_n^{\qa}G(x',x)
	(\pileft_j+\vec{\pi}_j)I_{-n}^{\qa}] \nn\\
	& = & (\fr{1}{2\me})^2
	\sum_k G_{k+n}(x,x')(\pileft_i+\vec{\pi}_i)G_k(x',x)
	(\pileft_j+\vec{\pi}_j)
\eea
The frequency sum can be split in two parts: (I) $k$ and $k\!+\!n$ have the
same sign; (II) $k$ and $k\!+\!n$ have opposite signs.
Case II has been done in great detail in the context of the SCBA. The
conclusion is that (II) does not contribute either to $\qs_{\rm xx}$ or
$\qs_{\rm xy}$ when $\mu$ is in a density of states gap. 
Case I for $i\!\neq\! j$, using the relation 
$\pileft\!+\!\vec{\pi}\!=\!-i2\me[G^{-1},\vec{x}]$, 
gives rise to the familiar `Streda' form for $\qs_{\rm xy}$.
For $i\!=\!j$, the last two contributions in (\ref{X2approx}) sum up to zero.
We arrive at the following expression,
\be
	X_2\approx \Tr\ln G^{-1}+\half m\sum_{n\qa}\intdxx
	n\; \vec{z}_n^{\qa}\times\vec{z}_{-n}^{\qa}.
\label{result2}
\ee
Now we have to find a match between the first order result (\ref{result1})
for $T\!\neq\! {\bf 1}$, $W\!\neq\! {\bf 1}$ and the second order result 
(\ref{result2}) for $T\!=\!{\bf 1}$, $W\!=\!{\bf 1}$.
Up to a constant arising from the difference between
$G_0$ and $G$, this match is given by
\bea
	&& \Tr\ln G^{-1}+\pi\qr_{\rm edge}\ointdx \tr\qo R
	+m\left(\fr{1}{8}\qe^{ij}\Tr R[D_i,R]
	[D_j,R]-\fr{i}{2}\Tr R\curl \hat{\vec z}
	\right) \nn\\
	&&=\Tr\ln G^{-1}+\fr{m}{2\vd}\oint\! dx\; \tr R(\qo-i\vd\hat{z}_x)
	+mS_{\rm top}[R]-\fr{im\qb}{4\pi}\intdxx\;
	\vec{z}\dagg\times\prt_0
	\vec{z} 
\label{match}
\eea
with $\vd$ the electron drift velocity at the edge,
\be
	\vd=m/(2\pi\qr_{\rm edge}).
\ee
Writing this result in terms of 
$ Q$ is nontrivial, since the ${\rm I}$-matrices appearing
in $W$ are truncated and do not satisfy a $U(1)$ algebra.
As a consequence of the truncation procedure, quadratic terms in 
${A}'$ arise.
Using the relations
\bea
	\tr\qo R & = & \tr\qo Q
	+\tr\hat{{A}_0'} Q-\fr{\qb}{2\pi}
	{{A}_0'}\dagg  {A}_0' \nn\\
	\tr R\hat{z}_x & = &
	\tr Q\hat{A}_x-\tr Q\prt_0\inv(\prt_x{A}_0')
	-\fr{\qb}{\pi}{A}_x\dagg {A}_0'
	+\fr{\qb}{\pi}{{A}_0'}\dagg \prt_0\inv(\prt_x{A}_0')
	\nn\\
	S_{\rm top}[R] &=& S_{\rm top}[Q]-\fr{i}{2}\ointdx
	\tr Q\prt_0\inv(\prt_x{A}_0')
	+\fr{i\qb}{4\pi}\ointdx
	{{A}_0'}\dagg \prt_0\inv(\prt_x{A}_0')
	\\
	\qb\intdxx \vec{z}\dagg\!\times\!\prt_0\vec z &=&
	-\qb\intdxx \qe^{\mu\nu\qk}(A_\mu')\dagg(\prt_\nu A_\qk') 
	-\qb\ointdx [\prt_0{A}_x-\prt_x{A}_0'] \dagg
	\prt_0\inv A_0' \nn
\eea
which are a result of the peculiar ``${\cal F}$-algebra'' structure
(\ref{I,I}), we obtain the following action
\bea
\label{afterWrot}
	S[ Q,{A},\ql] &=& S_{\rm c}[\ql]
	+S_{\rm b}[\ql,{A}]+S_{\rm Q}[Q,\ql,{A}]  \\
	S_{\rm c} &=& -\half\qb\intd{^2 x d^2 x'}
	\ql\dagg(x)U_0\inv(\vec x-\vec x')\ql(x') \nn\\
	S_{\rm b} &=& \fr{im\qb}{4\pi}
	\intdxx \qe^{\mu\nu\qk}(A_\mu')\dagg(\prt_\nu A_\qk')
	-\fr{m\qb}{4\pi\vd}\oint\! dx\;
	{{A}_0'}\dagg{A}_-' \nn\\
	S_{\rm Q} &=& \fr{m}{2\vd}\oint\! dx\; 
	\tr Q(\qo+\hat A_-')
	+mS_{\rm top}[Q]. \nn
\eea
The first term is the Coulomb energy contribution from the plasmon
field; the $S_{\rm b}$ is a ``boson'' action (this adjective will
become clear later on); the last expression, $S_{\rm Q}$, contains
the action for the $Q$ field and the coupling of $Q$ with 
$\ql$ and ${A}$.
We have defined a `minus' direction as follows,
\be
	{A}_-' = {A}_0'-i\vd{A}_x
\ee
reflecting the chirality inherent in the problem.

\ns{Chiral edge bosons}
\label{secchiral}

In this Chapter we take the theory one step further and derive the
theory of chiral edge bosons, similar to the one obtained by 
Wen\cite{bigWen}
in a phenomenological approach to abelian quantum Hall states.
For noninteracting electrons such a formulation is readily obtained
(Section~A). For interacting electrons, however, the procedure is more
complicated and we first derive an effective Finkelstein-type action
of the $Q$-field at the edge, obtained by eliminating the plasmon
field $\ql$ (Section~B1). In Section~B2 we show that the theory
provides complete information on the response of the system to
external fields. We derive an edge anomaly for the interacting
electron gas and show the connection with Laughlin's gauge
argument. The complete theory for interactions as well as the 2+1
dimensional Chern-Simons theory are derived in Section~B3. In
Section~C we give some explicit results on the single particle Green's
function which enters the expression for electron tunneling into the
quantum Hall edge. This, then, completes the theory of the integral
quantum Hall edge.

\subsection{The noninteracting case}
\label{secnoninteract}

In the case of free electrons, only the fields $ Q$ and ${A}$
are present.
In order to obtain an effective action for $A_{\mu}$ we integrate
out $Q$. We make use of (\ref{GminGplus}) with $2\qo\!\naar\!\qo_n$
and 
%
obtain the  tree level propagator
\be
	\left\langle \vphantom{M^M}\tr\Ian Q(-q)\; \tr\Iamn Q(q)
	\right\rangle
	=\fr{\qb\vd}{2\pi m}\fr{\qo_n}{\qo_n+i\vd q}.
\ee
This yields the result
\be
	S=\fr{im\qb}{4\pi}\left[
	\intdxx\qe^{\mu\nu\qk}A_\mu\dagg\prt_\nu A_\qk
	+\oint\! dx\; E_x\dagg\left(\prt_-\inv A_- +\qO\res\right)
	\right]
\label{free case}
\ee
with the following meaning of the symbols:
\be
	\prt_-=\prt_0-i\vd\prt_x
\ee
and the inverse $\prt_-\inv$ is given by
\bea
	(\prt_-\inv F)(x,\qt) &=& \fr{1}{(2\pi)^2}\intd{x' d\qt'}
	F(x',\qt')\int_{-\infty}^{\infty}\! dk\int_{-\infty}^{\infty}d\qo
	\; \fr{\exp[ik(x-x')-i\qo(\qt-\qt')]}{-i\qo+\vd k} \nn\\
	&=& 
	\fr{1}{2\pi}\intd{x' d\qt'}
	F(x',\qt')\left[\fr{\qy(\qt-\qt')}{\vd(\qt-\qt')-i(x-x')+\qh}
	+
	\fr{\qy(\qt'-\qt)}{\vd(\qt-\qt')-i(x-x')-\qh}\right]
\eea
with $F$ an arbitrary function,
$\qy$ the step function and
$\qh$ a regulator. The operation $\prt_-\inv$ does not
commute with $\prt_-$. On the one hand
it is easily checked that
$\prt_-(\prt_-\inv F)=F$, but on the other hand we have
\be
	\prt_-\inv(\prt_- F) = F-F\res 
\ee
with $F\res$ defined as that part of $F$ which satisfies $\prt_- F=0$.
Another property of this operation is
\be
	\intd{xd\qt}F_1(\prt_-\inv F_2)=
	-\intd{xd\qt}(\prt_-\inv F_1)F_2.
\ee
The $Q$-integration can also be done by choosing $\qO$ in such a way
that $Q$ decouples from $A_-$,
\bea
	\prt_-\qO=A_- \hskip1cm &;& \hskip1cm 
	\qO=\prt_-\inv A_- +\qO\res,
\label{choicephi}
\eea
yielding exactly the same result (\ref{free case}).
The action (\ref{free case})
can also be written as a path integral over $m$ charge 1 bosons,
\be
	S[A,\qf_i]=\fr{i\qb}{4\pi}
	\sum_{i=1}^{m}\left[
	\intdxx \qe^{\mu\nu\qk}A_\mu\dagg \prt_\nu A_\qk
	-\oint\! dx\;(D_x\qf_i\dagg D_-\qf_i-E_x\dagg \qf_i)
	\right],
\label{nonintphi}
\ee
where the covariant derivative $D$ is defined as
$D_\mu\qf_i\!=\!\prt_\mu\qf_i\!-\!A_\mu$.
The zero-momentum part of each boson field has to be excluded from the
path integral, since the action does not depend on it. 
In order to make contact with Ref.~1 we mention that
(\ref{nonintphi}) 
is equivalent to a Chern-Simons bulk theory with  $m$ 
gauge fields
$g^i$ that represent potentials for the electron currents, coupled to
the external potentials $A_\mu$, 
\be
	S[A,g^i]=\fr{i\qb}{4\pi}\sum_{i=1}^{m}
	\intdxx\qe^{\mu\nu\qk} \left[\vphantom{\sum}
	-(g^i_\mu)\dagg\prt_\nu g^i_\qk+2 (g^i_\mu)\dagg\prt_\nu A_\qk
	\right]
\label{coupling}
\ee
where the $g^i$ have the gauge fixing constraint 
$g^i_-|_{\rm edge}=0$.
In appendix B we explicitly show how integration over the potentials
$g^i$ leads to the action (\ref{nonintphi}).

\subsection{The Coulomb case}
\label{secCoulomb}
\subsubsection{Integration over $\ql$ and Q}

Now we look at the full action (\ref{afterWrot}). 
In this expression
the plasmon field $\ql$ is contained in the following way
\bea
\label{plasmoncontent}
	S_{\rm b}[{A}'] &=& S_{\rm b}[{A}]
	+\fr{im\qb}{2\pi}\intdxx\;\ql\dagg{B}
	-\fr{m\qb}{4\pi\vd}\oint\! dx\; (\ql\dagg\ql+2\ql\dagg{A}_0)
	\\ 
	S_{\rm Q}[{A}'] &=& S_{\rm Q}[{A}]
	+\fr{m}{2\vd}\oint\! dx\; \tr\hat{\ql} Q. \nn
\eea
Integrating out the plasmon field $\ql$, we obtain an effective action
for $Q$ coupled to ${A}$, which we organize as follows
\be
	S = S_0[Q]+S_{\rm int}[Q,{A}]+S_{\rm b}[{A}]
	+S_{\rm flux}[{A}].
\label{afterlambdaS}
\ee
The first term is given by
\be
	S_0[Q]= mS_{\rm top}[Q]+S_{\rm F}[Q]
	-\fr{m\pi}{4\qb}\sum_{n\qa}\int\! \fr{dk_x}{2\pi}\fr{1}{\veff(k_x)}
	|\tr \Ian Q|^2
\label{afterlambda0}
\ee
with 
\be
	S_{\rm F}[Q]=
	\fr{m\pi}{4\qb\vd}\ointdx \left[\sum_{n\qa} \tr\Ian Q \tr\Iamn Q 
	+4\tr\qo Q\right]
\ee
the edge analogue of the ${\cal F}$-invariant Finkelstein action for the
bulk [I] and
\be
	\veff(k_x) = \vd+mU_0(k_x)
\label{defveff}
\ee
the ``effective velocity'',
where $U_0(k_x)=(2\pi)^{-1}\intd{k_y}U_0(\vec k)$ is
the Coulomb interaction on the edge.
The last term in (\ref{afterlambda0}) is the edge version of the
``Coulomb'' term from [I]. Note that the Finkelstein and ``Coulomb''
terms together can be written as
\be
	\fr{m}{2\vd}\ointdx \tr\qo Q
	+\fr{m\pi}{4\qb\vd}\sum_{n\qa} \int\!\frac{dk_x}{2\pi}
	\frac{\qr_{\rm edge}}{U_0\inv(k_x)+\qr_{\rm edge}}
	|\tr\Ian Q|^2	
\ee
where the expression in front of the $|\tr {\rm I}Q|^2$ is just the 1D
screened Coulomb interaction.
The other terms in (\ref{afterlambdaS}) are a coupling term
\be
	S_{\rm int}[Q,{A}]= \fr{m}{2}\int\!\fr{dk_x}{2\pi}
	\fr{1}{\veff(k_x)}
	\tr Q\hat{A}_c\eff,
\label{afterlambdaint}
\ee
a ``boson'' term
\be
	S_{\rm b}[{A}] = 
	\fr{im\qb}{4\pi}\intdxx 
	\qe^{\mu\nu\qk}(A\eff_\mu)\dagg \prt_\nu A\eff_\qk
	-\fr{m\qb}{4\pi}\int\!\fr{dk_x}{2\pi}\fr{1}{\veff(k_x)}
	({A}_0\eff)\dagg{A}_c\eff,
\label{afterlambdab}
\ee
and a flux-flux interaction term
\be
	S_{\rm flux}[{A}] = 
	\fr{\qb}{2}(\fr{m}{2\pi})^2\intd{^2 x d^2 x'}
	{B}\dagg(\vec{x})U_0(\vec{x}-\vec{x}'){B}(\vec{x}').
\ee
Here we have introduced 
an `effective' gauge field which contains a Coulomb correction to the
scalar potential,
\bea
	\vec{A}^{\rm \eff} = \vec{A} \hskip5mm &;& \hskip5mm
	{A}_0^{\rm eff}(\vec{x})={A}_0(\vec{x})
	+\fr{im}{2\pi}\intd{^2 x'}
	U_0(\vec{x}-\vec{x}')B(\vec{x}'),
\label{defAeff}
\eea
and an effective `minus'-direction denoted by the subscript `c'
\bea
	\prt_c=\prt_0-iv\eff\prt_x \hskip1cm &;& \hskip1cm
	{A}_c={A}_0-iv\eff {A}_x.
\label{defprtc}
\eea
Comparing the result (\ref{afterlambdaS}))
with the free particle case
(\ref{afterWrot} without $\ql$) we see that 
the presence of the Coulomb interaction has the following effects:
\begin{itemize}
\item the appearance of the flux-flux interaction term 
$S_{\rm flux}[{A}]$ and of the
screened Coulomb interaction in $S_0[Q]$.
\item the replacements ${A}_0\!\naar\! {A}_0^{\rm eff}$
and ${A}_-\!\naar\! {A}_c^{\rm eff}$.
\item the replacement $\vd\!\naar\! v\eff(k_x)$.
\end{itemize}
For what follows, it is convenient to 
rewrite the first three terms of 
(\ref{afterlambdaS}) as
\bea
	S_0+S_{\rm int}+S_{\rm b} &=&
	\fr{im\qb}{4\pi}\left[
	\intdxx \qe^{\mu\nu\qk}(A\eff_\mu)\dagg\prt_\nu A\eff_\qk
	-\ointdx {A}_x\dagg 
	{A}_c\eff\right] + m\Stop[Q] \nn\\
	&& +S_{\rm F}[Q]
	-\fr{m\pi}{4\qb}\sum_{n\qa}\int\!\fr{dk_x}{2\pi}
	\fr{1}{\veff(k_x)}
	\left|\tr \Iamn Q-\fr{\qb}{\pi}({A}_c\eff)^\qa_n\right|^2
\label{complsq}
\eea
where, as in the bulk [I], the gauge field couples to $Q$ only via the
gauge non-invariant ``Coulomb'' term in (\ref{afterlambda0}). However,
compared to the bulk case where the coupling results in the gauge
invariant combination $(\tr{\rm I} Q-\fr{\qb}{\pi}A_0)$, the situation
is more subtle in the edge case. The expression 
$(\tr{\rm I}Q-\fr{\qb}{\pi}{A}_c\eff)$ appearing in (\ref{complsq})
is, in fact, gauge variant, but this gauge variance is exactly what
one needs to compensate for the edge contributions resulting from
gauge transformations of the ``boson'' action $S_{\rm b}$ and the topological
term. Therefore, the complete action (\ref{complsq}+$S_{\rm flux}$) is
fully gauge invariant.

We now proceed as in section~\ref{secnoninteract} and
integrate out the $Q$ field.
This is done in the same way as for the
noninteracting case; either by doing it directly or by choosing 
$\qO$ such that $Q$
decouples from ${A}_\mu$, i.e.
\be
	\prt_c\qO=A^{\rm eff}_c.
\ee
The only difference lies in the fact that we now work with effective
quantities.
The arguments about the `residual' part of the electric field can
again be applied, but now for the effective quantities 
(\ref{defveff}, \ref{defAeff}, \ref{defprtc}).
We then get the effective action for the external field $A_\mu$ in the
presence of Coulomb interactions,
\bea
\label{Coulombcase1}
	S[A] &=& 
	\fr{im\qb}{4\pi}\left[
	\intdxx \qe^{\mu\nu\qk}(A\eff_\mu)\dagg\prt_\nu A\eff_\qk
	+\oint\! dx\;
	(\prt_c\inv A_c\eff +\qO\res)\dagg E_x\eff\right]
	\\ & &
	+\fr{\qb}{2}(\fr{m}{2\pi})^2\intd{^2 x d^2 x'}
	B\dagg(\vec{x})U_0(\vec{x}-\vec{x}')B(\vec{x}'). \nn
\eea
Again, the difference with the free particle case is the appearance of
a flux-flux term and various replacements by effective quantities.

\subsubsection{Edge currents and Laughlin's gauge argument}
\label{secargument}
The action (\ref{Coulombcase1}) contains complete information on the
response of the system to external electromagnetic fields.
We define the current as 
$j^\mu(\vec x)\!=\!\qd S/\qd A_\mu(\vec x)$. In this way we find
\bea
\label{currentj0}
	j^0(\vec x)&=& \fr{im}{2\pi}\left[B-
	\qd(y)\prt_c\inv E_x\eff\right] \\
	j^1(\vec x)&=& -\fr{im}{2\pi}\left[E_y-\prt_y 
	\intd{^2 x'}U_0(\vec x-\vec x')j^0(\vec x')\right]
	-\fr{m\vd}{2\pi}\qd(y)
	\prt_c\inv E_x\eff \\
	j^2(\vec x) &=& \fr{im}{2\pi}\left[E_x-
	\prt_x \intd{^2 x'}U_0(\vec x-\vec x')j^0(\vec x')\right].
\label{currentj2}
\eea
It is easily verified that $\prt_\mu j^\mu\!=\!0$.
The edge currents are obtained by taking only those terms
that possess a delta function. On the edge we get
\bea
	j^0_{\rm edge} &=& -\fr{im}{2\pi}
	\prt_c\inv E_x\eff
	\\
	j^1_{\rm edge} &=& -i\vd\cdot j^0_{\rm edge}.
\eea
This yields for the edge anomaly
\be
	\prt_\mu j^\mu_{\rm edge}(x)=-\fr{im}{2\pi}
	\left[E_x-\prt_x\intd{^2 x'}U_0(x,\vec x')j^0(\vec x')\right].
\label{anomaly}
\ee
By applying Laughlin's gauge argument\cite{Prange} one can now
directly relate the 
conductances defined by the bulk and by the edge. 
For example, let us do a linear response calculation for
the case where $N$ flux quanta $h/e$ are created somewhere inside a hole in 
the sample. 
The charge $q$ flowing from one edge into the other is found
using (\ref{anomaly}),
\be
	dq/d\qt=-i\oint \prt_\mu j^\mu_{\rm edge}=
	\fr{m}{2\pi} d\qF/d\qt
\ee
where $\qF$ is the total flux $N\cdot h/e$ enclosed by the contour integral.
This yields $q\!=\!m\cdot N$, as it should.

\subsubsection{Interacting chiral bosons}
As was the case in the free electron situation,
we can write the theory (\ref{Coulombcase1})
as an edge boson coupled to the external
field, exactly of the form (\ref{nonintphi}),
but now with effective quantities and an extra flux-flux term,
\bea
	S[A,\qf_i] &=& \fr{i\qb}{4\pi}
	\sum_{i=1}^{m}\left[
	\intdxx \qe^{\mu\nu\qk}(A\eff_\mu)\dagg\prt_\nu A\eff_\qk
	-\oint\! dx\;(D_x\qf_i\dagg D_c\eff\qf_i- \qf_i\dagg E_x\eff)
	\right]\nn\\
	&& +\fr{\qb}{2}(\fr{m}{2\pi})^2\intd{^2 x d^2 x'}
	B\dagg(\vec{x})U_0(\vec{x}-\vec{x}')B(\vec{x}').
\label{bosonsCoulomb}
\eea
As in the noninteracting case, this result is equivalent to a
Chern-Simons bulk theory of the form (\ref{coupling}).
In this case the action for the electron currents is given by
\bea
	S[A,g_i] &=& \fr{i\qb}{4\pi}\sum_{i=1}^{m}
	\intdxx\qe^{\mu\nu\qk} \left[\vphantom{\sum}
	-(g^i_\mu)\dagg\prt_\nu g^i_\qk
	+2 (g^i_\mu)\dagg\prt_\nu A\eff_\qk\right] \nn\\
	& & +\fr{\qb}{2}(\fr{m}{2\pi})^2\intd{^2 x d^2 x'}
	B\dagg(\vec{x})U_0(\vec{x}-\vec{x}')B(\vec{x}')
\label{coupling2}
\eea
with the gauge fixing  conditions 
\be
        \left[g^j_-(k_x)-i\fr{m}{\sqrt{2\pi}}U_0(k_x)\sum_{a=1}^{m}
        g^a_x(k_x)\right]_{\rm edge}=0.
\ee
It is very instructive to write (\ref{bosonsCoulomb}) 
also in the following way
\bea
\label{Cbosonform}
	S &=& 
	-\fr{\qb}{2}(\fr{1}{2\pi})^2\sum_{i,j=1}^{m}\intd{^2 x d^2 x'}
	U_0(\vec{x}-\vec{x}')\curl[\qy(y)\vec{D}\qf_i(\vec x)]\dagg
	\nabla'\!\times\![\qy(y')\vec{D}\qf_j(\vec x')]
	\nn\\
	&& +\fr{i\qb}{4\pi}
	\sum_{i=1}^{m}\left[
	\intdxx \qe^{\mu\nu\qk}A_\mu\dagg\prt_\nu A_\qk
	-\oint\! dx\;(D_x\qf_i\dagg D_-\qf_i-E_x\dagg \qf_i)
	\right].
\eea
Notice that 
there are no effective quantities in this expression;
the Coulomb interaction is completely contained in the first term.
The charge density is given by
$\fr{m}{2\pi}[B\!+\!\qd(y)m\inv\sum_i D_x\qf_i]$.
Notice also that we have written a two-dimensional integral containing
$\qf_i$, even though the boson fields only exist on the edge. This is
not a problem, since the $\qf_i$ only get evaluated at the edge.

\subsection{Tunneling density of states}
\label{sectunnel}
In [I] we expressed the one particle Green's function which enters the
tunneling density of states in terms of the matrix $Q$ variable as follows
\be
	\langle Q^{\qa\qa}(\qt_2,\qt_1,\vec{x}_0) \rangle =
	\sum_{n=-\infty}^{\infty}e^{i\nu_n(\qt_2-\qt_1)}
	\langle Q_{nn}^{\qa\qa}(\vec{x}_0)\rangle.
\ee
The gauge transformation that in (\ref{complsq})
decouples $Q$ from ${A}$
introduces into the path integral over (\ref{Coulombcase1})
an extra factor
\be
	\exp -i\left([\prt_c\inv A_c\eff]^\qa(\qt_2,\vec{x}_0)
	-[\prt_c\inv A_c\eff]^\qa(\qt_1,\vec{x}_0)\right).
\ee
When decoupling the quadratic edge term in $A$ (\ref{Coulombcase1})
with the use of boson fields, this factor translates to 
\be
	\exp -i\int_{\qt_1}^{\qt_2}\!\!\! 
	d\qt\;\prt_\qt\qf_j^\qa(\qt,\vec x_0),
	\;\;\;\;\; j=1,\ldots,m
\label{expphi}
\ee
in (\ref{bosonsCoulomb}).
The decoupling is not a unique procedure, since combinations of
the boson fields $\qf_i$ can be chosen other than (\ref{expphi}).
However, the above form is the only one that 
yields the fermionic exponent for the expectation value
$\langle Q\rangle$.
\be
	\left\langle
	\exp -i\int_{\qt_1}^{\qt_2}\!\!\! d\qt\;\prt_\qt\qf_i^\qa(\qt,\vec x_0)
	\right\rangle\propto (\qt_2-\qt_1)^{-S}
	\;\;\;\;\;\;\;\;  S=1.
\label{S=1}
\ee
(See appendix A for the explicit calculation.)
Notice that we would have had a serious problem at this point if we had not
excluded the zero-momentum components of the $\qf_i^\qa$ when we introduced
these auxiliary fields. A redefinition of
the integration measure, 
$\int\!\!{\cal D}\qf\!\naar\!\int\!\!{\cal D}[\qf\!+\!f]$, 
with $\prt_x f(x,\qt)\!=\!0$, 
would yield a result depending on the arbitrary function $f$.

\ns{Long-range disorder}
\label{seclongrange}

In Section~\ref{secplatrev} we introduced the idea of percolating edge
states as a model for smooth, slowly varying randomness. Application
of $Q$-field theory then provides an effective and elegant way of
describing the transport properties of the network model near the
percolation threshold. In this Chapter we extend the network theory of
percolating edge states in several ways. We show that the Coulomb
interactions can dramatically alter the behavior of the electron gas,
depending on the physical process that one is interested in. The
concept of a `tunneling density of states', that describes the
tunneling of electrons into the quantum Hall edge, is particularly
sensitive to the presence of long range electron-electron
interactions.
In Sections~A--C we derive an `effective' theory of chiral edge bosons
that includes the effect of Coulomb interactions between the edge and
bulk electrons. This leads to a tunneling exponent $S$ that varies
continuously with the filling fraction $\nu$ like $1/\nu$. This result
is in dramatic contrast to the Fermi liquid predictions of
Section~\ref{sectunnel} which apply to isolated edges alone. We start
out (Section~A) with the chiral boson formulation of the network model
and employ the Laughlin gauge argument in order to illustrate the
fundamental differences between transport and edge tunneling
(Section~B). 
Section~C describes one of the most important aspects of this
Chapter. It deals with the detailed mechanism by which the `neutral'
modes are eliminated from the effective theory for edge tunneling.
We end this Chapter with a computation of the inelastic relaxation
rate (Section~D) that enters into the transport problem at finite
temperatures (Section~\ref{secintereff}).

\begin{figure}
\begin{center}
\setlength{\unitlength}{1mm}
\begin{picture}(80,70)
\put(0,0)
{\epsfxsize=80mm{\epsffile{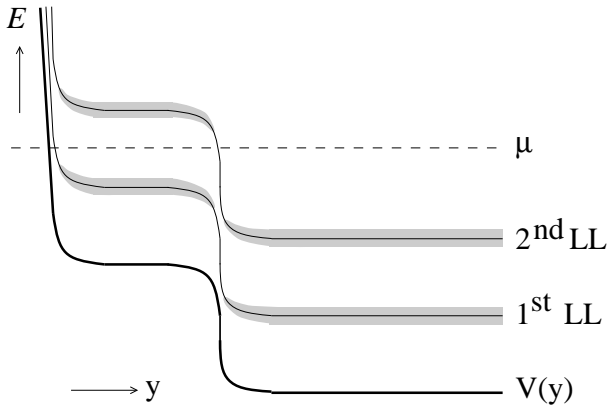}}}
\end{picture}
\caption{Spatially separated  edge channels}
\label{channelsplit}
\end{center}
\end{figure}

\subsection{Separation of edge channels}

Long-range disorder can cause the edge states of different Landau
levels to become spatially separated. A potential fluctuation at the edge
can lift all states in such a way that new `edge' states are created.
(See Fig.~\ref{channelsplit}) If the chemical potential lies between the
shifted and unshifted energy of a Landau level, the edge states of this
Landau level will be situated inside the sample, not on the outermost edge.
If there are several potential jumps of this kind, all the edge channels
can become separated.
They can also start wandering into the interior of the sample.

We propose that `edge channel separation' is the dominant effect of
smooth potential fluctuations as opposed to `inter-channel scattering'
which only occurs when the potential changes abruptly. In this section
we wish to embark on the problem of smooth potential fluctuations in
the presence of the Coulomb interactions.

\begin{figure}
\begin{center}
\setlength{\unitlength}{1mm}
\begin{picture}(140,50)
\put(0,5)
{\epsfxsize=140mm{\epsffile{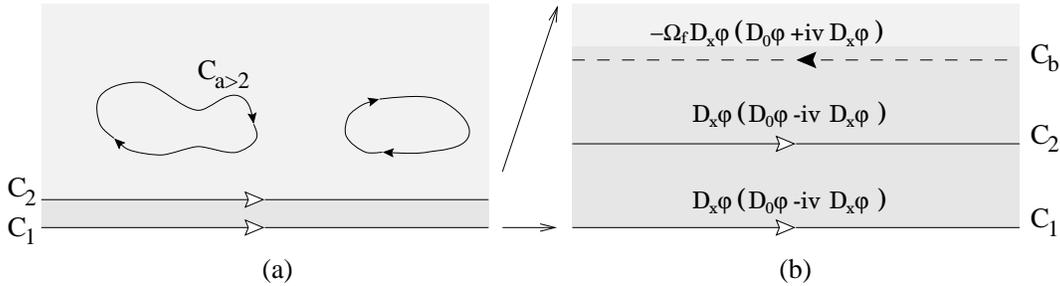}}}
\end{picture}
\caption{(a) Plot of equipotential contours corresponding to filling fraction 
$\nu\! = \!2 \! -\!\qe$. (b) Effective edge theory for filling fraction
$\nu\! = \!2 \! -\!\qe$. The dashed line represents the (anti)chiral
contribution from the bulk orbitals.}
\label{figislands}
\end{center}
\end{figure}

In order to fix the thought we imagine a quantum Hall sample with
filling fraction 
$\nu\! = \!2 \! -\!\qe$. 
Fig.~\ref{figislands}a illustrates the
equipotential contours. We may distinguish between the localized
(closed) orbitals in the bulk of the sample and the extended (chiral)
edge states.

This picture leads us to the idea of describing the chiral bosons by {\em
one} field $\qf(\vec x)$ that lives on all the `edges' instead of
independent fields for every edge.
The action (\ref{bosonsCoulomb}) then becomes
\bea
	S&=&\fr{i\qb}{4\pi}\left[
	\intdxx n(\vec x)\qe^{\mu\nu\qk}A_\mu\dagg\prt_\nu A_\qk
	-\sum_{a=1}^M s_a
	\oint_{C_a}\!\!\! dx\left(D_x\qf\dagg [D_0\qf-is_a\vd D_x\qf]
	-E_x\dagg\qf\right) \right] \nn\\
	&& -\fr{\qb/2}{(2\pi)^2}\intd{^2 xd^2 x'}U_0(\vec x-\vec x')
	\curl[n(\vec x)\vec{D}\qf(\vec x)]\dagg\nabla'\!\times\!
	[n(\vec x')\vec{D}\qf(\vec x')].
\label{splitS[phi,A]}
\eea
The $n$ is a function of position labeling the `local' filling
fraction: 
outside the sample $n(\vec x)$ 
is zero; going inward, it increases by one
every time you cross an edge, until it
reaches its bulk value $m$.
At the bulk orbitals, $n(\vec x)$ jumps again. (In the case 
$\nu\! =\! 2\! -\!\qe$, depicted in Fig.~\ref{figislands}a, 
$n(\vec x)\!=\!1$ inside
the closed orbitals.)

Each edge is described by a contour labeled $C_a$, with 
$a\!=\!1,\cdots,m$
for the edge states and $a\!=\!m\!+\!1,\cdots,M$ for the closed bulk orbitals.
The coordinate `$x$', appearing in the edge terms, is defined on the
contour and is taken in the positive (anticlockwise) direction.
The symbol $s_a$,
\be
	s=(\overbrace{+1,\cdots,+1}^m,-1,\cdots,-1)
\ee
incorporates the fact that the contours with
$a\! \leq\! m$ and $a\! > \! m$ 
carry opposite current and charge densities.
For simplicity we take the drift velocity
$\vd$ the same for all edges.
Integrating out the boson field yields the generalization of 
(\ref{Coulombcase1}),
\bea
\label{splitCoulombcase}
	S[A] &=& 
	\fr{i\qb}{4\pi}\left[
	\intdxx n(\vec x) \qe^{\mu\nu\qk}(A\eff_\mu)\dagg\prt_\nu A\eff_\qk
	+\sum_a s_a \oint_{C_a}\!\!\! 
	dx\; [\prt_c\inv(\prt_x A_c\eff)
	-A_x]\dagg A_c\eff
	\right]
	\\ & &
	+\fr{\qb}{2}(\fr{1}{2\pi})^2\intd{^2 x d^2 x'}
	n(\vec x)B\dagg(\vec{x})U_0(\vec{x}-\vec{x}')n(\vec x')B(\vec{x}'). \nn
\eea
The notation $\prt_c$ (at contour $C_a$) now has the sign $s_a$ in front
of the velocity and contains
Coulomb interactions with {\em all}
contours instead of just $C_a$ itself.
The definition of the `effective' potential $A_0\eff$ has also slightly
changed, 
\be
	A_0\eff(\vec x)=A_0(\vec x)+\fr{i}{2\pi}
	\intd{^2 x'}U_0(\vec x-\vec x')\; n(\vec x') B(\vec x').
\ee
For completeness, in appendix D we also present the generalization of
the action $S[Q,{A}]$ 
(\ref{afterlambdaS})
for the case of separated edge channels.
Note that we are addressing the situation where the chemical
potential is away from the narrow `percolation' regime indicated by
$W_0$ in Fig.~\ref{figdos}. We will next exploit the simplicity of our
model and demonstrate that the Hall conductance and the tunneling
density of edge states are fundamentally different quantities that
correspond to completely different physical processes.

\subsection{Hall conductance}
\label{secLaughlin2}
First, it is straightforward to generalize the results of 
section~\ref{secargument} to include the separated edge channels and
the bulk states into Laughlin's flux argument. 
Differentiating the action (\ref{splitCoulombcase}) with respect to $A_\mu$,
we obtain the generalized form of the currents
(\ref{currentj0}--\ref{currentj2}),
\bea
\label{splitj0}
	j^0(\vec x)&=& \fr{i}{2\pi}\left[n(\vec x)B-
	\sum_{a=1}^M s_a\qd(\vec x \;{\rm on}\; C_a)
	\prt_c\inv E_x\eff\right] \\
	j^i(\vec x)&=& -i\fr{n(\vec x)}{2\pi}\qe^{ij}\left[E_j-\prt_j 
	\intd{^2 x'}U_0(\vec x-\vec x')j^0(\vec x')\right]
	-\fr{\vd}{2\pi}\sum_{a=1}^M \qd(\vec x \;{\rm on}\; C_a)
	\prt_c\inv E_x\eff (\vec e^{\;a}_\parallel)_i
\label{splitjvec}
\eea
where 
the vector $\vec e^{\;a}_\parallel$ is tangent to the contour $C_a$ and 
points in the positive direction.
Again it is easy to check that $\prt_\mu j^\mu\!=\!0$, i.e. that current
conservation is respected. The edge currents are given by
\bea
	j^0_{\rm edge}(C_a) &=& -\fr{i}{2\pi}s_a 
	\prt_c\inv E_x\eff
	\\
	j^x_{\rm edge}(C_a) &=& -is_a\vd \cdot j^0_{\rm edge}(C_a).
\eea
The edge anomaly applies to each bulk orbital and edge
state separately,
\be
	\prt_\mu j^\mu_{\rm edge}(C_a)=-\fr{i}{2\pi}s_a
	\left[E_x-\prt_x\intd{^2 x'}U_0(x,\vec x')j^0(\vec x')\right].
\label{splitanomaly}
\ee
As expected, the sign $s_a$ determines whether charge is
transported into an edge or from an edge into the bulk.
By repeating Laughlin's flux argument it is now
demonstrated explicitly that the localized bulk orbitals do not affect
the transport of charge from one sample edge to the other, independent of the
electron-electron interactions; taking (\ref{splitanomaly}) and performing
the contour integral over $C_a$ we obtain the charge transported per unit of
time from the $a$'th channel,
\be
	dQ_a/d\qt=-i\oint_{C_a}\!\!dx\; \prt_\mu j^\mu_{\rm edge}
	=s_a\fr{1}{2\pi} d\qF_a/d\qt
\ee
where $\qF_a$ is the magnetic flux enclosed by $C_a$.
For $a \! >\! m$ this flux is obviously zero, 
since the localized bulk orbitals do not encircle the hole in the sample.
This, then, shows that the Hall conductance is quantized (equal to $m$) 
independent of $\qe$.

\subsection{Tunneling density of states}
Laughlin's flux argument for the Hall conductance
expresses the quantum Hall state as an exact
`excited' state of the system. Tunneling processes into the edge, on
the other hand, are expressed in terms of eigenstates near the Fermi
energy, i.e. the tunneling density of states, and 
due to the Coulomb interactions
this quantity is
sensitive to the presence of bulk orbitals.
We start from the action (\ref{splitS[phi,A]}), omitting the replica
indices for notational simplicity and putting $A_\mu\!=\!0$,
\bea
	S &=& -\fr{i}{4\pi}\intd{\qt}\!\!\sum_{j=1}^Ms_j\oint_{C_j}\! 
	dx\; \prt_x\qf(\prt_0\qf-is_j\vd\prt_x\qf) \nn\\
	&& -\fr{1}{8\pi^2}\intd{\qt}\!\!\sum_{j,j'=1}^M s_j s_{j'}
	\oint_{C_j}\!dx\oint_{C_{j'}}\!\!dx'\;
	\prt_x\qf U_0(x,x')\prt_{x'}\qf.
\label{Mislands}
\eea
Following section~\ref{sectunnel}, (\ref{expphi}), the
one particle Green's function can be written as follows
\be
	G(\qt_2-\qt_1)=\left\langle\vphantom{H^H}
	\exp -i[\qf(\qt_2,x_0)-\qf(\qt_1,x_0)] \right\rangle
\label{Greensfun}
\ee
where $x_0$ denotes a point on the edge contour $C_1$. 
The presence of the Coulomb interactions makes the computation of $G$
a complicated two dimensional problem. Some procedure
needs to be found which extracts the lowest energy excitations from
(\ref{Mislands}). We follow the strategy of taking the boson fields as
a two dimensional field variable and we then collect the terms with
smallest momenta. This procedure is done in position space and we
proceed by giving the details of a step by step analysis.
The results for the tunneling exponents are given 
at the end of~C4,
which also contains a brief summary in the end.

\subsubsection{Gradient expansion}
\label{secgradientexp}
The interaction term
in (\ref{Mislands}) can be written as a sum over area integrals,
\be
	-\fr{1}{8\pi^2}\intd{\qt}\!\!\sum_{j,j'=1}^M\int_{C_j}\!\! d^2 x
	\!\int_{C_{j'}}\!\! d^2 x'\; s_j s_{j'}\;
	\prt_a\qf(\vec x) U_{ab}(\vec x-\vec x')\prt_b\qf(\vec x')
\label{areaU}
\ee
with
\be
	U_{ab}(\vec x-\vec x')=
	\qe_{ac}\qe_{bd}\fr{\prt}{\prt x_c}\fr{\prt}{\prt x_d'}
	U_0(\vec x-\vec x').
\ee
Since we are only interested
in the $\qf$ with the smallest momenta, we can make the replacement
\be
	\sum_{j=m+1}^M\int_{C_j}\! d^2 x\; \naar \qO_f\int_{C_b}\! d^2 x.
\label{smearintegral}
\ee
The $\qO_f$ stands for the fraction of the total
area that is enclosed by all the bulk orbitals together. 
The contour $C_b$ is not sharply defined and is located somewhere
close to the edge (see Fig.~\ref{figislands}b).
It encloses the region within which the bulk orbitals are contained.
The joint Coulomb effects of the bulk orbitals will effectively be comprised on
this contour.
For the terms in (\ref{Mislands}) containing $\prt_x\qf\prt_0\qf$ we can write
\be
	\sum_{j>m}\oint_{C_j}\!\!\!dx\; \prt_x\qf\prt_0\qf =
	\sum_{j>m}\int_{C_j}\!\!\! d^2x\; \curl(\nabla\qf \prt_0\qf)
	\naar \qO_f\!\!\int_{C_b}\!\!\! d^2 x\; \curl(\nabla\qf \prt_0\qf)
	=\qO_f\!\!\oint_{C_b}\!\!\!dx\; \prt_x\qf\prt_0\qf.
\ee
The expression 
${\displaystyle \sum_{j>m}\oint_{C_j}\!\!\!dx\; (\prt_x\qf)^2}$
averages out to 
${\displaystyle\qk\int_{C_b}\!\!\! d^2 x\; (\nabla\qf)^2}$ 
with $\qk$ some positive
constant related to the total length of all the bulk contours.
If there are substantial stretches where a bulk orbital runs along the
edge, interaction terms will arise, leading to
a term $\oint_{C_b}\!dx\; (\prt_x\qf)^2$.

Note that in doing the replacement (\ref{smearintegral})
in
(\ref{areaU}), one also needs to introduce correction terms that compensate
for the errors made when the separation 
$|\vec x\!-\!\vec x'|$ is `small'
(of the order of the average size of the orbitals or less)
and $U_{ab}$ does not vary slowly. These corrections are of the form
$\intdxx (\nabla\qf)^2$.

Then there are also extra correction terms that will arise 
if there are regions where a bulk orbital runs along the
edge. This correction takes the form of a short-ranged interaction between
$C_b$ and all the other contours (including $C_b$).

Having done the replacement (\ref{smearintegral}) and writing
the interaction terms again as contour integrals, we have the following
action, 
\bea
\label{aftersmearing}
	S &=& -\fr{i}{4\pi}\int\! d\qt\left[\sum_{j=1}^m\oint_{C_j}\!\!
	dx\;\prt_x\qf\prt_-\qf-\qe\oint_{C_b}\!\!\! dx\;
	\prt_x\qf\prt_0\qf
	\right] \\ &&
	-\fr{1}{8\pi^2}\int\! d\qt \left[\sum_{jj'=1}^m\oint_{C_j}\!\!\!
	dx\oint_{C_{j'}}\!\!\!\! dx'\; \prt_x\qf U \prt_{x'}\qf
	+\qe^2\oint_{C_b}\!\!\! dxdx'\;\prt_x\qf (U+V_b) \prt_{x'}\qf
	\right. \nn\\ && \left. 
	-2\qe\sum_{j=1}^m\oint_{C_j}\!\!\! dx\oint_{C_b}\!\!\! dx'\;
	\prt_x\qf (U+V_j) \prt_{x'}\qf \right]
	-g\int\!\! d\qt\!\! \int_{C_b}\!\!\! d^2 x\; (\nabla\qf)^2,
	\nn
\eea
where $g$ is a positive constant. We have identified $\qO_f$ with
$\qe$, since the fraction of the area occupied by bulk states is
exactly the deviation from integer filling.
We have written $V_b(x,x')$ for the short-ranged interaction between two
points on $C_b$; The $V_j(x,x')$ denotes the short-ranged interaction
between a point $x$ on $C_j$ and $x'$ on
$C_b$. 
The precise expression for $V$ is unknown due to the fact that it has
its origin in the twilight zone near the edge, where it is unclear
whether a term contributes to the bulk or edge action.

Comparing this result (\ref{aftersmearing}) 
with (\ref{Mislands}), we see that the presence
of the interacting bulk states effectively leads to the appearance of
an additional (anti)chiral boson on the
contour $C_b$,
an extra short-ranged interaction with this contour,
and a
lower dimensional left over bulk term $\int\! (\nabla\qf)^2$.

\subsubsection{Effect of the bulk term}
\label{secbulkterm}

In order to be able to calculate the tunneling density of states 
(\ref{Greensfun}) we need an effective theory for the edge degrees of
freedom, and therefore we have to understand how they are affected by the left
over bulk term. To this end, we are going to split bulk and edge degrees of
freedom.
We write the bulk term as $\int_{C_b}\!d^2 x (\nabla\qF)^2$, where
$\qF$ represents 
the bulk degrees of freedom and is treated as an integration variable
independent of $\qf$. To reflect the fact that it is actually an extension of
$\qf$ into the bulk, we impose some boundary condition on $\qF$, for instance
$\qF|_{\rm edge}=\qf$ or $\prt_\perp\qF|_{\rm edge}=\prt_\perp\qf$.
($\prt_\perp$ is the derivative perpendicular to the contour.)
The effect of the bulk term on the edge theory is obtained by integrating out
$\qF$, which leads to an effective action for the boundary conditions.
Let us consider a general scenario and impose the boundary conditions
$\qF|_{\rm edge}=\qj_0$ and $\prt_\perp\qF|_{\rm edge}=\qj_1$, using
constraint multipliers $k_0$ and $k_1$, respectively.
\bea
\label{phiwithconstr}
	e^{S_{\rm eff}[\qj_0(x),\qj_1(x)]}
	&=& \int\!\!{\cal D}[\qF(\vec x)]\;{\cal D}[k_0(x)]\;
	{\cal D}[k_1(x)] \times \\ & \times &
	\exp\left\{i\ointdx k_0(\qF-\qj_0)
	+i\ointdx k_1(\prt_\perp\qF-\qj_1)-g\intdxx (\nabla\qF)^2
	\right\}. \nn
\eea
For notational simplicity we have omitted time dependence and the subscript
$C_b$ under all the integrals.
We first wish to integrate (\ref{phiwithconstr}) over $\qF(\vec x)$ keeping
$k_0$ and $k_1$ fixed. For this purpose we split $\qF$, which has free
boundary values,
into a bulk and an edge part by writing
\bea
	\qF=\qF_L+\hat\qF \hskip.5cm & 
	\prt_\perp\qF_L|_{\rm edge}=\prt_\perp\qF|_{\rm edge} \hskip.5cm&
	\prt_\perp\hat\qF |_{\rm edge}=0
\label{PhiLPhihat}
\eea
where $\qF_L$ satisfies Laplace's equation
\be
	\nabla^2\qF_L(\vec x)=0.
\ee
The $\qF_L(\vec x)$ is completely determined by
$\prt_\perp\qF_L$ on the edge, which we now take as an independent edge
degree of freedom denoted by $E_1(x)$.
Introducing the 2D Green's function $G$,
\bea
	G(\vec x,\vec x')=\fr{1}{2\pi}\ln|\vec x-\vec x'| \hskip1cm &;& \hskip1cm
	\nabla^2 G(\vec x,\vec x')=\qd(\vec x-\vec x'),
\label{Groen}
\eea
and using Green's theorem,
we solve Laplace's equation and obtain for $\qF_L(\vec x)$
\be
	\qF_L(\vec x)=-\oint\! dx'\; \left[G(\vec x,x')E_1(x')
	-\qF_L(x')\fr{\prt G}{\prt y'}(x,y; x',0)\right].
\ee
This expression tells us that we need to now
$\qF_L$ on the edge in order to evaluate $\qF_L$ in the bulk. 
Luckily, we do not need the full 2D $\vec{x}$ dependence, since due
to the splitting (\ref{PhiLPhihat}) $\qF_L$ will get evaluated at the
edge only.
Using a special property of the Green's function (\ref{Groen}),
namely $[\prt_{y'}G](x,0; x',0)\!=\!0$,
we can explicitly write $\qF_L$ on the edge as a function of $E_1$,
\be
	\qF_L(x)=-\oint\!\! dx'\; G(x,x')E_1(x').
\ee
The action, written in terms of $\hat\qF$ and $E_1$, is now given by
\bea
	S &=& -g\intdxx(\nabla\hat\qF)^2-g\oint\!\oint\! E_1 G E_1
	+2g\ointdx E_1\hat\qF    -i\oint\!\oint\! Gk_0 \nn\\ &&
	+i\ointdx k_0(\hat\qF-\qj_0)+i\ointdx k_1(E_1-\qj_1)
\label{splitPhiE1}
\eea
where we have used the shorthand notation $\oint\!\oint\! AGB$ for the
expression  \linebreak
$\oint \! dx\!\oint\! dx' A(x)G(x,x')B(x')$.
Integrating out $\hat\qF$ is now simply done by replacing $\hat\qF$ by its
saddle point value.
Varying the action with respect to $\hat\qF$, keeping $E_1$ fixed, we
get the saddlepoint equation
\be
	\nabla^2\hat\qF+\qd(y)[E_1+\fr{i}{2g}k_0]=0.
\ee
Using the Green's function's property
$[\prt_{y'}G](x,0; x',0)\!=\!0$ again,
we find the following solution on the edge
\be
	\hat\qF(x)=-\oint\!\! dx' \; G(x,x')
	[E_1+\fr{i}{2g}k_0](x').
\label{Phihatedge}
\ee
In substituting this solution into (\ref{splitPhiE1}) 
we do not need the full 2D $\vec x$-dependence of $\hat\qF(\vec x)$, since we
can write 
$\intdxx(\nabla\hat\qF)^2\!=\!-\intdxx\hat\qF\nabla^2\hat\qF$
and $\nabla^2\hat\qF$ is an expression restricted to the edge.
Substitution of (\ref{Phihatedge}) into (\ref{splitPhiE1}) yields
\be
	S = -2g\oint\!\oint\! E_1 G E_1
	+\fr{1}{4g}\oint\!\oint\!  k_0 G k_0 
	 -i\ointdx k_0 \qj_0    -2i \oint\!\oint\! G E_1
	+i\ointdx k_1(E_1-\qj_1).
\ee
Integrating out $k_0$ is straightforward and gives
\be
	S = g\oint\!\oint\! \left(\qj_0 G\inv \qj_0
	+2 E_1 G E_1\right) +4g\ointdx \qj_0 E_1
	+i\ointdx k_1(E_1-\qj_1).
\ee
In the end we integrate out $k_1$, yielding the constraint 
$E_1\!=\!\qj_1$.
The final result for $S_{\rm eff}[\qj_0,\qj_1]$ becomes 
\bea
	S_{\rm eff}[\qj_0,\qj_1] &=& g\oint\!\oint\!
	\left(\qj_0 G\inv\qj_0+2\qj_1 G\qj_1\right) 
	+4g\ointdx \qj_0\qj_1
	\nn\\  &=&
	g\oint\!\oint(\qj_0, \qj_1)\left(
	\matrix{G\inv & 2 \cr  2 & 2G }\right)
	\left(\matrix{\qj_0\cr \qj_1}\right).
\label{Spsipsi}
\eea
We are going to put $\qj_0\!=\!0$ in order to avoid double counting of 
$(\prt_x\qf)^2$ terms at the edge, and $\qj_1\!=\!\prt_\perp\qf$. 
The action (\ref{Spsipsi}) becomes
\be
	S[\prt_\perp\qf]=2g\oint\!\oint\! \prt_\perp\qf \; G\; \prt_\perp\qf.
\label{Sperpphi}
\ee
This edge term, derived from the interaction with the bulk orbitals,
is seriously going to affect the tunneling exponent.
A quick way to see this is as follows: on the contours 
$C_1,\cdots, C_b$, the field $\qf(\vec x)$ can be written as 
$\qf(x,y\; {\rm on }\; C_1)+$ perpendicular derivatives. For the
tunneling exponent, only $\qf|_{C_1}$ is needed, so we can integrate
out the perpendicular derivatives in 
(\ref{aftersmearing} minus bulk term+\ref{Sperpphi})
to obtain an effective action for
$\qf$ on $C_1$. The dominant part of the 1D propagator for $\prt_\perp\qf$
is given by $G\inv(k)\!\propto\! |k|$, from which it follows that all terms
introduced by the integration over $\prt_\perp\qf$ are irrelevant. 
Higher powers of $\prt_\perp$ are even less relevant.
Replacing all the $\qf$ in (\ref{aftersmearing}) by $\qf|_{C_1}$, we
get a term $\nu\ointdx \prt_x\qf\prt_0\qf$, leading to a tunneling
exponent $S\!=\!1/\nu$ instead of the free particle result $S\!=\!1$.

In the next section we are going to derive this result more formally,
based on a consideration of the neutral modes in the theory where
the edge channels are not spatially separated.

\subsubsection{Demise of the neutral modes; example $\nu\!=\!1\!-\!\qe$}
\label{demise}

In the long wavelength limit, the contours $C_1,\cdots,C_b$ are lying
so close together that we can effectively return to the picture where
all the edge channels are sitting on top of each other.
We label the channels $\qf_1(x),\cdots,\qf_m(x), \qf_b(x)$.
Let us for simplicity's sake first consider the case 
$\nu\!=\!1\!-\!\qe$, where
we just have the two fields $\qf_1$ and $\qf_b$.
In terms of these fields, the action (\ref{aftersmearing}), 
without the bulk term and the bulk effect (\ref{Sperpphi}), takes the
form (again using abbreviated notation)
\bea
\label{labelform1}
	S_0[\qf_1,\qf_b] &=& -\fr{1}{4\pi}\ointdx [\prt_x\qf_1\prt_0\qf_1
	-\qe \prt_x\qf_b\prt_0\qf_b ] \\
	&& -\fr{1}{8\pi^2}\oint\!\oint \!U 
	[\prt_x\qf_1-\qe\prt_x\qf_b]^2
	-\fr{1}{8\pi^2}\sum_{k,l=1,b}
	\ointdx V_{kl}\prt_x\qf_k\prt_x\qf_l. \nn
\eea
We have put all the short-range contributions into the 
$2\!\times\!2$ velocity matrix $V$.
We next define a `charged mode' $\qG$ and a `neutral mode' $\qg$ in
such a way that only the charged mode feels the long-range part of the
interaction,
\bea
	\qG=\fr{1}{\nu}(\qf_1-\qe\qf_b) \hskip1cm &;& \hskip1cm
	\qg=\qf_1-\qf_b \nn\\
	\qf_1=\qG-\fr{\qe}{\nu}\qg
	\hskip1cm &;& \hskip1cm 
	\qf_b=\qG-\fr{1}{\nu}\qg.
\eea
In the basis ($\qG,\qg$) the action (\ref{labelform1}) becomes
\bea
	S_0[\qG,\qg] &=& -\fr{1}{4\pi}\ointdx\left[\nu\prt_x\qG\prt_0\qG
	-\fr{\qe}{\nu}\prt_x\qg\prt_0\qg\right] 
	 -\fr{\nu^2}{8\pi^2}\oint\!\oint\! U (\prt_x\qG)^2 \nn\\
	&& -\fr{1}{8\pi^2}\ointdx [\prt_x\qG \;\; \prt_x\qg ]
	\hat V
	\left[\matrix{\prt_x\qG \cr \prt_x\qg}\right]
\eea
where $\hat V$ is the velocity matrix in the new basis.
The expression $\prt_\perp\qf$ in the theory for spatially separated
channels is in the single-edge picture evidently equivalent to the neutral mode
$\qg\!\propto\!\qf_b\!-\!\qf_1$. The leftover bulk contribution 
(\ref{Sperpphi}) therefore translates into an extra term involving the
neutral mode,
\be
	S_{\rm bulk}[\qg]={\rm const.}\cdot\oint\!\oint\! \qg \; G\; \qg.
\label{Sbulkgamma}
\ee
The tunneling density of states is now expressed as
\be
	\left\langle \exp -i\qf_1 |_{\qt_1}^{\qt_2}\right\rangle
	\propto
	\int\!{\cal D}[\qG]{\cal D}[\qg]\;
	\exp\left(\vphantom{H^H}
	-i(\qG-\fr{\qe}{\nu}\qg)|_{\qt_1}^{\qt_2}+
	S_0[\qG,\qg]+S_{\rm bulk}[\qg]\right).
\ee
If we perform the integration over $\qg$ first, 
we see that the `bulk' part of the action yields the following
contribution to the  inverse propagator:
$G(k)\!\propto\! 1/|k|$, which is dominant at low momenta.
The integration over $\qg$ yields $\qG$-$\qG$ terms of order
$k^5 \hat V(k)$. These are clearly irrelevant. For the tunneling
density of states we can write
\bea
	\left\langle \exp -i\qf_1 |_{\qt_1}^{\qt_2}\right\rangle
	& \propto & \int\!{\cal D}[\qG]
	\exp \left(\vphantom{H^H} -i\qG|^{\qt_2}_{\qt_1}
	+S_{\rm eff}[\qG]\right) \nn\\
	S_{\rm eff}[\qG] & = &
	-\fr{i\nu}{4\pi}\ointdx\prt_x\qG\prt_0\qG
	-\fr{\nu^2}{8\pi^2}\oint\!\oint \prt_x\qG U\prt_x\qG
	-\fr{1}{8\pi^2}\ointdx \prt_x\qG \hat V_{\qG\qG}\prt_x\qG.
\label{Scharged}
\eea
For small momenta the $\hat V$ essentially 
reduces to a constant and we can use the
results of appendix~A, obtaining
\bea
	\left\langle \exp -i\qf_1 |_{\qt_1}^{\qt_2}\right\rangle
	\propto (\qt_2-\qt_1)^{-S} \hskip1cm &;& \hskip1cm
	S=1/\nu.
\label{oneovernu}
\eea

\subsubsection{The general case $\nu\!=\!m\!-\!\qe$}
The results for $\nu\! =\! 1 \! -\! \qe$ are easily generalized.
From the `bulk' channel $\qf_b$ and the edge channels 
$\qf_1,\cdots, \qf_m$ we construct a charged mode $\qg_0$ and $m$
neutral modes $\qg_1,\cdots,\qg_m$ as follows,
\bea
	\qg_0 &=& \fr{1}{\nu}(\sum_{k=1}^m\qf_k-\qe\qf_b) \nn\\
	\qg_a &=& \fr{1}{a}(\sum_{k=1}^a\qf_k-a\qf_{a+1})
	\;\;\;\;\;\;\;  a=1,\cdots,m
\eea
where we define $\qf_{m+1}$ as $\qf_b$. 
The neutral modes $\qg_1,\cdots,\qg_{m-1}$ 
are the usual ones for a theory with
$m$ edges. They are mutually perpendicular and normal to the charged
mode.
The additional $\qg_m$ is normal to the other neutral modes but not to
the charged mode.
The $\qf$'s are expressed in
terms of the $\qg$'s as follows
\bea
	\qf_b &=& \qg_0-\fr{m}{\nu}\qg_m \nn \\
	\qf_k &=& \qg_0-\fr{\qe}{\nu}\qg_m-(1-\fr{1}{k})\qg_{k-1}
	+\sum_{a=k}^{m-1}\fr{1}{a+1}\qg_a \;\;\;\;\; k\leq m.
\eea
Equation (\ref{labelform1}) is generalized to
\bea
\label{labelform}
	S[\qf] &=& -\fr{1}{4\pi}\ointdx\left[\sum_{j=1}^m\prt_x\qf_j\prt_0\qf_j
	-\qe \prt_x\qf_b\prt_0\qf_b \right] \\
	&& -\fr{1}{8\pi^2}\oint \!\oint \! U 
	\left[\sum_{j=1}^m\prt_x\qf_j-\qe\prt_x\qf_b\right]^2
	-\fr{1}{8\pi^2}\sum_{k,l=1}^{m+1}
	\ointdx V_{kl}\prt_x\qf_k\prt_x\qf_l. \nn
\eea
Again, all the short-range contributions have been put into a 
velocity matrix $V$, which now has dimension 
$(m\! +\! 1)\!\times\!(m\! +\! 1)$. 
Writing (\ref{labelform}) in terms of the $\qg$-basis, we get
\bea
	S[\qg] &=& -\fr{1}{4\pi}\ointdx\left[\nu\prt_x\qg_0\prt_0\qg_0
	+\sum_{a=1}^{m-1}\fr{a}{a+1}\prt_x\qg_a\prt_o\qg_a
	-m\fr{\qe}{\nu}\prt_x\qg_m\prt_0\qg_m\right] \nn\\
	&& -\fr{\nu^2}{8\pi^2}\oint\!\oint\! U (\prt_x\qg_0)^2
	-\fr{1}{8\pi^2}\sum_{a,c=0}^m\ointdx 
	\hat V_{ac}\prt_x\qg_a\prt_x\qg_c
\eea
where $\hat V$ is the velocity matrix in the basis of $\qg$'s.
The argument of 
(\ref{Sbulkgamma} to \ref{oneovernu}) can be applied again, in a slightly
modified form; the neutral modes are equivalent to $\prt_\perp\qf$ and
higher derivatives. (A basis $\hat\qg$ can be found for the neutral
modes in which 
$\hat\qg_n$ corresponds to the 1D lattice discretization of 
$\prt_\perp^n\qf$.) 
On dimensional grounds
the propagator for the $n$'th
normal derivative of $\qf$ has to be proportional to $k^{2n-1}$, leading to
irrelevant contributions. 
A more concrete way of making this statement would be to generalize the
analysis presented in (\ref{phiwithconstr} to \ref{Spsipsi}),
including boundary conditions for the higher normal derivatives.
However, that would also require us to take into account 
higher order terms in the
$\qf$-theory (\ref{aftersmearing}).
The resulting effective action for the charged mode $\qg_0$ is of the form
(\ref{Scharged}), with $\nu\! =\! m\! -\!\qe$.

We can summarize the results of section~\ref{seclongrange} 
as follows: We have seen that the Fermi liquid result $S\!=\!1$ is
obtained for the tunneling density of states (i) when the
Coulomb interactions are omitted, or (ii) when interactions are included but 
only short length scales are considered.
An interacting theory for the lowest lying excitations, which are
slowly varying field configurations,  yields completely different
results. The presence of bulk orbitals, interacting mutually and with
the edge states, is effectively described by an extra edge channel 
with prefactor $-\qe$
plus
a remnant of the interactions in the bulk of the form
$\int(\nabla\qf)^2$. The leftover bulk term serves to make all the
neutral edge modes irrelevant, yielding an effective edge action for
the one remaining, charged, mode. Due to the presence 
of the extra `bulk' channel, the prefactor of this effective action $S[\qG]$
becomes $m \! -\! \qe\! =\! \nu$, which is a continuous parameter in
sharp contrast to the integer quantized $m$.
For the tunneling exponent we obtain $S\!=\!1/\nu$.

\subsection{Computation of $\tau_{\rm in}$}
\label{sectauin}
We next return to the problem of the plateau transitions. Following
section~\ref{secplatrev} we expect that the transport at high temperatures
is dominated by interactions between the conducting
electrons on the backbone saddlepoint network and those on the
disconnected pieces or clusters.

\begin{figure}
\begin{center}
\setlength{\unitlength}{1mm}
\begin{picture}(80,80)(0,0)
\put(0,0)
{\epsfxsize=80mm{\epsffile{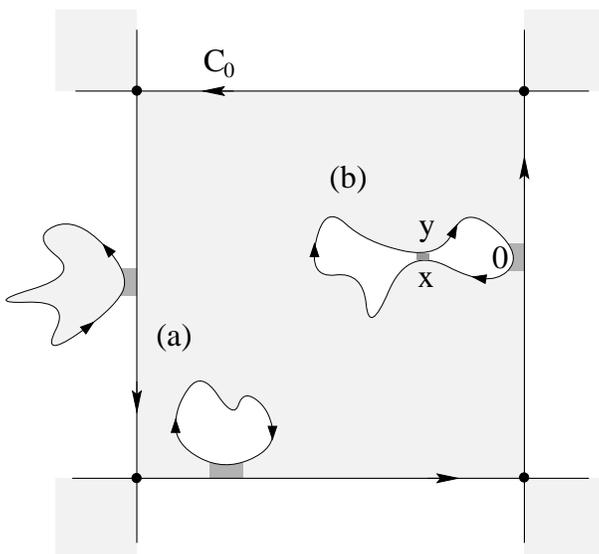}}}
\end{picture}
\caption{(a) Point-like interaction between conducting electrons 
and the localized electrons at 'nearly saddlepoints'. (b) The coordinates
$0$, $x$ and $y$ are the points of interaction along the localized 
contour.}
\label{figCoulomb}
\end{center}
\end{figure}

The fundamental
quantity to compute is the characteristic time $\qt_{\rm in}$ that is
needed for the backbone electrons to equilibrate with the rest of the
network. In order to set up a theory for relaxation, we consider the
`nearly saddlepoints' in the network, where tunneling is not possible
but where the Coulomb forces nevertheless produce `sudden changes' in
the motion of the conducting electrons.
Fig.~\ref{figCoulomb} illustrates the interaction of the saddlepoint
network with disconnected orbitals. The `nearly saddlepoints' where
the Coulomb forces are most effective are
indicated by the shaded areas. We can model the situation by
introducing a delta-function potential which acts in the small areas
of the nearly saddlepoints only. The action can be written as
\be
	S_{\rm eff}[\qf]=S[\qf_0]+\sum_i S[\qf_i]
	-\sum_i\intd{\qt}\prt_x\qf_0(\vec a_i) U_i \prt_x\qf_i(\vec a_i)
\label{C_1}
\ee
where $S[\qf_0]$ is the action for the chiral boson field on a link of
the saddlepoint network that we denote as the contour $C_0$,
\be
	S[\qf_0]=\intd{\qt}\oint_{C_0}\!\!\! dx\; \prt_x\qf_0\prt_-\qf_0.
\ee
This contour is taken to be very large or infinite. Similarly, we
define chiral boson fields $\qf_i$ on the disconnected but large
contours $C_i$,
\be
	S[\qf_i]=-\intd{\qt}\oint_{C_i}\!\!\! dx\; \prt_x\qf_i
	\prt_+\qf_i.
\ee
The sum in the interaction term in (\ref{C_1}) is over the discrete
set of nearly saddlepoints $\vec a_i$ along the contour $C_0$ where
the fields $\qf_0$ and $\qf_i$ interact with an appropriate, random
strength $U_i$. This problem is in many ways quite similar to the
problem of interacting edge channels with a randomly varying
separation between them. We proceed along the same lines as in
\cite{Finkelstein} and introduce a self energy $\qS$ for the
density-density correlation of the field $\qf_0$.
If we denote the Fourier transforms of the propagators
$\langle\prt_x\qf_0(x,\qt)\prt_x\qf_0(x',\qt')\rangle$ and
 $\langle\prt_x\qf_i(x,\qt)\prt_x\qf_i(x',\qt')\rangle$
(with $x,x'$ parametrizing the positions on the contours $C_0$, $C_j$
respectively) as
\bea
	D_0(\qo,q)=\frac{-iq}{i\qo-\vd q} \hskip0.5cm &;& \hskip0.5cm
	D_j(\qo,q)=\frac{-iq}{i\qo+\vd q},
\eea
then the introduction of a self energy takes the form
\be
	D_0(\qo,q)\naar -i\frac{q}{i\qo-(\vd+\qS)q}.
\ee
To lowest order in the interaction potential we may write
\be
	\qS(\qo)=-iz\overline{U_j^2}\int\!\fr{dq}{2\pi}D_j(\qo,q)=
	\fr{z}{2\vd^2} \overline{U_j^2} |\qo|.
\label{Selfenergy}
\ee
Here, the bar stands for the average over the random positions $\vec
a_i$ along $C_0$ and $z$ is the linear density of saddlepoints.
The result (\ref{Selfenergy}) can be used to obtain an expression for
$1/\qt_{\rm in}$, i.e. the imaginary part of the self energy as it
appears in the electron Green's function ${\cal G}(\qo,q)$ as follows
\be
	1/\qt_{\rm in}=\int\!\fr{\qo dq}{(2\pi)^2}\qS(\qo){\cal G}(\qe-\qo,q).
\ee
The $\qt_{\rm in}$ determines the rate at which the electrons on the
backbone cluster equilibrate with the rest of the electronic orbitals.
We find $\qt_{\rm in}\inv\propto \qe^2$ or $T^2$ at finite
temperatures. This admittedly crude approach toward electron
relaxation can be improved in several ways. For example, as the most
important correction to the self energy (\ref{Selfenergy}) we find the
self-interacting orbitals as depicted in Fig.~\ref{figCoulomb}b.
These corrections replace the momentum integral in (\ref{Selfenergy})
in the following way (in space-time notation)
\bea
	\int\!\fr{dq}{2\pi}D_j(\qo,q) &=& \intd{\qt} e^{-i\qo(\qt-\qt')}
	D_j(0,0; \qt-\qt') \nn\\
	D_j(0,0; \qt-\qt') &\naar & D_j(0,0; \qt-\qt') \nn\\ &&
	+ \intd{\qt_0}\!\!\int_0^L\!\! dx\!\!\int_x^L\!\!\! dy\;
	D_j(0,x;\qt-\qt_0)\; \tilde U_j\; D_j(y,L; \qt_0-\qt')
\label{replaceD}
\eea
where $x,y$ are the positions of the nearly saddlepoint where the
self-interaction takes place. 
The integrals stand for the averaging over positions and all
dimensional factors are absorbed into $\tilde U_j$.
The length of the orbital is given by $L$ and boundary conditions
$x \!\equiv\! x\!+\!L$ and $y\! \equiv\! y\!+\!L$ are understood.
Equation (\ref{replaceD}) can be rewritten as a shift in the chemical
potential,
\be
	\intd{q}D_j(\qo,q)\naar \intd{q}
	\frac{-iq}{i\qo-\qd\mu+\vd q}\;\;\;\;\;\;\;\;\;\; 
	\qd\mu=\tilde U_j.
\ee
This leads to a modified self
energy according to
\be
	\qS(\qo)\naar \fr{z}{2\vd^2} \overline{U_i^2}(\qo+i\qd\mu)
	\sgn(\qo).
\ee
The shift $\qd\mu$ can be translated into a shift in the expression
for $\qt_{\rm in}\inv$ following
\be
	\qt_{\rm in}\inv(\qe,\qd\mu)=
	\left(1+i\qd\mu\fr{\prt}{\prt\qe}\right)\qt_{\rm in}\inv(\qe).
\ee
After the analytic continuation to real energies ($i\qe\naar \qe$)
has been performed, we obtain the final result
$\qt_{\rm in}\inv\propto\qe$ or $\qt_{\rm in}\inv\propto T$
at finite temperatures. More generally, we expect the equilibration
rate to be given by a regular series expansion in powers of $T$ which
is dominated by the lowest order
$\qt_{\rm in}\inv\!\propto\! T$ as $T$ approaches absolute zero.

\ns{Summary and conclusions}
\label{secsummary}

We have shown that massless edge excitations are an integral part of
the instanton vacuum theory with free boundary conditions. Massless
edge excitations have fundamental consequences for the `stability' of
topological quantum numbers and for the quantization of the Hall
conductances in particular.
We have
used the formalism of ${\cal F}$-algebra, introduced in our previous work,
and derived a complete theory of the edge. We have established the
fundamental connection between the instanton vacuum and Chern-Simons
gauge theory. Both theories have previously been studied
independently and with different physical objectives. 
We have shown that our new approach to edge physics enables one to
address several longstanding problems of smooth disorder and
interaction effects. We have pointed out that fundamental differences exist
between tunneling at the edge and electron transport. 
Transport experiments inject electrons directly into edge states; 
these electrons do not get enough time to equilibrate with the rest of
the sample and are therefore effectively decoupled from the bulk.
A tunneling measurement, however,  probes eigenstates of the whole
system, which 
involve not only edge electrons, but also localized bulk orbitals. 
Since tunneling
processes do not probe the incompressibility of the electron gas, they
are generally treated incorrectly by the theory of isolated edges.
By taking into account the effect of Coulomb interactions between the edge and
the localized bulk states, we have derived an effective edge theory
that predicts a tunneling exponent $1/\nu$.

For the plateau transitions we have constructed a percolation model of
interacting edges. We have shown how inelastic scattering at the
`nearly saddlepoints' sets the temperature scale at which the
transport coefficients cross over from mean field behavior to
critical scaling. 
This crossover can involve arbitrarily low temperatures and it
explains the `lack of scaling' in the transport data taken from
samples with long-range disorder at finite temperatures.
Our mean field expression for the conductances agrees with recent
empirical fits to transport data at plateau transitions.

The results of this paper serve as the basic starting point for a
subsequent one\cite{fracedge}
where we extend the theory to include the statistical
gauge fields and the fractional quantum Hall regime.

\vskip1cm

\noindent
{\bf ACKNOWLEDGEMENTS}\newline
This research was supported in part by
INTAS (grant \#96-0580).

\vskip1cm
\scez\renewcommand{\theequation}{A\arabic{equation}}
\noindent
{\Large\bf Appendix A: One-dimensional propagator with Cou\-lomb interaction}
\addcontentsline{toc}{section}{Appendix A}

\vskip0.4cm
\noindent
In this appendix we calculate the correlation function
$G(\qt,0)$ for the charged boson fields  $\qf_i$ (\ref{bosonsCoulomb}),
\be
	G(\qt,x)= \langle\qf_i(\qt,x)\qf_i(0,0)\rangle \;\;\;\;\;\;\; \qt>0.
\ee
In momentum and frequency space this correlator is given by
(we omit the label $i$ since it is of no consequence)
\be
	\langle \qf_a(k)\qf_{-b}(-k')\rangle=\frac{2\pi i}{\qb}
	\frac{\qd_{ab}\qd(k-k')}{k[\qo_a+ik v\eff(k)]}.
\label{propkomeg}
\ee
We write the Coulomb interaction and the effective velocity 
$v\eff$ in the following form
\bea
	U_0(k)=-c\sqrt{2\pi}\ln(k/\qL)^2 \hskip1cm &;& \hskip1cm
	v\eff(k)=-m c \ln(k/\qL D)^2
\eea
where $c$ is a positive constant indicating the strength of the
Coulomb interaction,
$\qL$ is an ultraviolet cutoff and $D\!=\!\exp(\vd/2m c)$.
We will only consider low momenta $|k|\!<\!\ql\qL$, with
$\ql \!<\! 1$, so that we are well away from the point where the
Hamiltonian  becomes negative.

We take the Fourier transform of (\ref{propkomeg}) and change the
frequency sum to an integral, writing 
$\sum_n\!\naar\!\fr{\qb}{2\pi}\intd{\qo}$,
\bea
	\prt_\qt G(\qt,0)&=&\frac{i}{2\pi}\int_{-\ql\qL}^{\ql\qL} \!\!dk 
	\; v\eff(k)
	\int_{-\infty}^\infty\frac{e^{i\qo\qt}d\qo}{\qo+ikv\eff(k)}
	\nn\\
	& = & -\int_{-\ql\qL}^{\ql\qL} 
	\!\!dk \; v\eff(k)\qy(-k v\eff) e^{k v\eff(k) \qt}.
\label{Gtheta}
\eea
The step function $\qy(-k v\eff)$ constrains
the integration interval to $k\!<\!0$. 
We can split the last expression in
(\ref{Gtheta}) into two parts, using
$\ln k\; dk\!=\!d(k\ln k\!-\!k)$, and get 
\be
	\prt_\qt G(\qt,0)=-\fr{1}{\qt}[1-(\fr{\ql}{D})^{2mc\qt\ql\qL}]
	-2mc\qL D\int_0^{\ql/D}\! du\;\exp[2mc\qt\qL D\cdot u\ln u].
\label{Gmpos}
\ee
The function $u\ln u$ is negative on the whole interval $(0,\ql/D)$,
since $\ql/D \!<\! 1$. If we now send the cutoff $\qL$ to infinity, the term
with the integral in (\ref{Gmpos}) will go to zero as $1/\ln\qL$. 
The term $(\ql/D)^{2mc\qt\ql\qL}$ also
vanishes, yielding the free particle result
\be
	G(\qt,0)=-\ln\qt+{\rm constant}.
\label{Gfree}
\ee

\vskip1cm
\scez\renewcommand{\theequation}{B\arabic{equation}}
\noindent
{\Large\bf Appendix B: Chern-Simons action for bulk currents}
\addcontentsline{toc}{section}{Appendix B}

\vskip0.4cm
\noindent
In this appendix we show that (\ref{nonintphi}) is equivalent to the
following  bulk action:
\be
	S[A,g^i]=\fr{i\qb}{4\pi}\sum_{i=1}^{m}
	\intdxx\qe^{\mu\nu\qk} \left[\vphantom{\sum}
	-(g^i_\mu)\dagg\prt_\nu g^i_\qk+2 (g^i_\mu)\dagg\prt_\nu A_\qk
	\right]
\label{appCSbulk}
\ee
with the condition $g^i_-=0$ on the edge.

The $g^i$ are 2+1 dimensional potentials from which the electron
current density $j$ for every Landau level can be found,
\be
	j_i^\mu\propto \qe^{\mu\nu\ql}\prt_\nu g^i_\ql.
\ee
Notice three important subtleties:
\begin{itemize}
\item
The coupling of $g$ with the electromagnetic gauge
field is of the form $\qe^{\mu\nu\qk}g_\mu\prt_\nu A_\qk$ 
instead of the expected
$\qe^{\mu\nu\qk}A_\mu\prt_\nu g_\qk\propto j_\mu A^\mu$. 
These expressions differ by an edge term. The second form is {\em not}
invariant under the gauge transformations 
$A_\mu\!\naar\! A_\mu\!+\!\prt_\mu\chi$; 
the expression $\qe^{\mu\nu\qk}\prt_\nu A_\qk$ 
on the other hand is manifestly gauge invariant.
\item 
Putting an arbitrary spacetime component of $g$ zero on the edge
ensures that the action is invariant under
$g_\mu\!\naar\! g_\mu\!+\!\prt_\mu\qk$, a gauge transformation that does not
affect the current 
density. Without such a condition, gauge invariance is broken at
the edge.
\item
Because of the invariance under 
$g_\mu\!\naar\! g_\mu\!+\!\prt_\mu\qk$,
a gauge fixing condition has to be specified for the path integration
over $g$, for instance  the Coulomb gauge 
$\nabla\cdot\vec g\!=\!0$.
\end{itemize}
Let us now for simplicity drop the replica indices $\qa$ and the Landau
level index $i$ (effectively setting $m\!=\!1$). 
Having taken the condition
$g_-|_{\rm edge}=0$,
the component $g_-$ in (\ref{appCSbulk}) multiplies the following
constraint: 
\be
	\curl (\vec g-\vec A)=0.
\label{appconstr}
\ee
After integration over $g_-$, what remains of the action is
\be
	\fr{i}{4\pi}\int\!d\qt\intdxx \left(-\vec g\times\prt_- \vec g
	+2\vec g\times [\nabla A_- -\prt_- \vec A]\right)
\label{appremnant}
\ee
subject to the constraint (\ref{appconstr}).
The general solution of (\ref{appconstr}) is given by
\be
	\vec g=\vec A-\nabla\qf
\ee
with $\qf(\vec x)$ a real scalar field which is now the only integration
variable that is left. Substitution into (\ref{appremnant}) yields an
action where $\qf$ features only on the edge,
\be
	S[\qf,A]=
	\fr{i}{4\pi}\intd{\qt}\left[
	\intdxx\qe^{\mu\nu\qk}A_\mu\prt_\nu A_\qk
	-\ointdx (D_x\qf D_-\qf	- \qf E_x)\right].
\label{appresult}
\ee
This is exactly of the form (\ref{nonintphi}).

One may worry that the path integration over $\qf$ is ill-defined,
because of the bulk degrees of freedom of $\qf$, which do not appear
in (\ref{appresult}). However, $\qf$ inherits something from the
gauge fixing condition of $g$. This is most easily seen in the case of
the Coulomb gauge; here,
$\qf$ has to satisfy $\nabla^2\qf\!=\!0$. 
This means that the bulk degrees of freedom
are completely determined by $\qf(x)$ at the edge 
(the well known case of Laplace's equation with Dirichlet boundary
conditions)
and therefore aren't independent integration variables.  

One final remark on the boundary condition $g_-\!=\!0$:
The Hamiltonian (density) corresponding to (\ref{appresult}) is given
by
$\vd(D_x\qf)^2$.
It is not allowed to choose a velocity $\vd\!<\!0$, since this
would lead to energies that are unbounded from below.
In general, the boundary condition has to be taken in such a way that
the velocity of the chiral bosons has the same sign as the prefactor
multiplying $\fr{i}{4\pi}$
in (\ref{appCSbulk}), otherwise the integration over $g$ is ill-defined on
the edge.

\vskip1cm
\scez\renewcommand{\theequation}{C\arabic{equation}}
\noindent
{\Large\bf Appendix C: Inter-channel scattering at the edge}
\addcontentsline{toc}{section}{Appendix C}

\vskip0.4cm
\noindent
In this appendix we describe the various steps of the standard
$Q$-field approach to (edge) disorder. For the general case of $m$
chiral edge channels, one can differentiate between different types of
disorder, depending on whether one allows inter-channel scattering or
not. Although the different scattering potentials do not give rise to
fundamentally different physical results, it is nevertheless important
to define the `effective' edge Hamiltonian (\ref{Hjj}) which gives
rise to the same result (\ref{Ssplit}) that was previously obtained
for 2D electrons. Below we shall show that the following $m$ channel
model satisfies our requirements
\be
	{\cal H}_{\rm edge}^{kk'}=-i\vd\qd_{kk'}\prt_x+V_{kk'}(x)
\label{Hkk}
\ee
where $V$ is a hermitian random matrix and the elements $V_{kk'}$ are
distributed with a Gaussian weight
\be
	P[V]=\exp\{ -\fr{1}{g}\ointdx \tr V^2\}.
\ee
The indices $k,k' \! =\! 1,\cdots,m$ label the edge channels.
The form (\ref{Hkk}) implies that single potential scattering, as
described by the 2D Hamiltonian
\be
	{\cal H}_{\rm 2D}=\fr{1}{2m_e}(\vec p-\vec A)^2+V(\vec x),
\label{H2D}
\ee
does not naively translate into single potential scattering for the
edge states as obtained by solving (\ref{H2D}) in the presence of an
edge (infinite potential wall). Rather than that, one should allow for
inter-channel scattering of the `pure' eigenstates as in (\ref{Hkk})
in order to reproduce the effect of dirt in the general 2D problem
(\ref{H2D}).
We start from the following generating function for the averaged free
particle propagators
\be
	Z=\int\!\!{\cal D}[\bar{\qj}\qj]\int\!\!{\cal D}[V] P[V]
	\exp \qb\sum_{p=\pm,\qa,jj'}\ointdx \bar{\qj}_p^{\qa,j}
	\left[(\mu+ip\qo)\qd_{jj'}-{\cal H}\edge^{jj'}
	\right]\qj_p^{\qa,j'}.
\label{appmultipsi}
\ee
Integration over randomness and introduction of the matrix field 
$\tQ^{\qa\qb}_{pp'}(x)$ by performing the Hubbard-Stratonovich trick
leads to
\be
	Z=\int\!\!{\cal D}[\tQ]\exp\left\{-\fr{1}{g}\Tr \tQ^2+m\Tr\ln
	[\mu+i\vd\prt_x+i\tQ+i\qo\qL] \right\}.
\label{apptrln}
\ee
Notice that the edge channel label is not present in the new field
variable $\tQ$, but it is simply contained in an overall factor $m$.
Notice also that the type of randomness as considered here has
previously been introduced in a different context by the name of
$N$-orbital scattering, where $N$ (here $m$) is commonly used for
saddle point and large-$N$ expansion purposes.

We will next make use of the simple analytic properties of our 1D
Hamiltonian and show that the saddlepoint technique yields, in fact,
exact results for all $m$ and that therefore there is no need to rely
on $m$ to be `large'. The stationary point equation for $\tQ$,
\be
	i[\tQ_{\rm sp}]^{\qa\qb}_{pp'}=\qd^{\qa\qb}\qd_{pp'}
	[e_0+(-1)^p i/2\qt],
\ee  
can be written as
\be
	\fr{2}{g}(e_0\pm i/2\qt)= -m\int_{-\infty}^\infty\! \fr{dq}{2\pi}
	\; [\mu-\vd q+e_0\pm i(1/2\qt+\qo)]\inv
	=\pm im/2\vd
\ee
with the simple solution $e_0\!=\!0$, $\qt\!=\!2\vd/(mg)$.
One may next replace the original $\tQ$-field by the following change of
variables,
\be
	\tQ\naar T\inv PT\naar \fr{1}{2\qt} T\inv\qL T
	= \fr{1}{2\qt}Q.
\label{Qchange}
\ee
Here, the 
$T\in\frac{SU(2N)}{S(U(N)\times U(N))}$ 
are unitary rotations and the block-diagonal hermitian 
$P^{\qa\qb}_{pp'}\!=\!\qd_{pp'}P^{\qa\qb}_p$ represent the longitudinal
components. Replacing $P$ by its saddlepoint value, as written in
(\ref{Qchange}), turns out to be an exact statement, 
valid for all $m$.
The reason is contained in the fact that the fluctuations in $P$ are
weighted by propagators with poles in either the positive or negative
imaginary momentum plane. All the momentum integrals therefore sum up
to zero, giving rise to a zero weight to all orders in the
$P$-fluctuations. The replacement 
of (\ref{Qchange}) is exact when
inserted in the $\Tr\ln$. 
Equation (\ref{apptrln}) factorizes into
\bea
\label{factorize}
	Z &=& Z_P\cdot Z_T \\
	Z_P &=& \int\!\!{\cal D}[P]\; I[P]\exp\{-\fr{1}{g}\Tr P^2\}
	\nn\\
	Z_T &=& \int\!\!{\cal D}[T]\;\exp\{m\Tr\ln[\mu+i\vd\prt_x
	+\fr{i}{2\qt}\qL+iB] \} \nn
\eea
where all $T$-dependence is contained in the quantity $B$ according to
\be
	B=\vd T\prt_x T\inv+\qo T\qL T\inv = \vd T D_0 T\inv .
\ee
Equation (\ref{factorize}) can be evaluated further, and to lowest few
orders in an expansion in $B$ we obtain an effective action which can
be written as
\bea
	Z_T &=& \int\!\!{\cal D}[T]\; \exp S_{\rm eff}[T] \nn\\
\label{appcomm1}
	S_{\rm eff}[T] &=& \fr{m}{2\vd}\Tr \qL B(x)-\fr{m\qt}{8\vd}
	\Tr[B(x),\qL]^2 +\cdots \\
	&=& \fr{m}{2}\ointdx \tr \qL T\prt_x T\inv
	+ \fr{m}{2\vd}\qo \ointdx \tr \qL Q 
	-\fr{m\qt\vd}{8}\ointdx \tr [D_0, Q ]^2.
\label{appcomm2}
\eea
The coefficients appearing in (\ref{appcomm2}) all have a clear
physical significance in the context of disordered edge states (see
also the main text). In particular, $m$ stands for the quantized Hall
conductance $\qs_{xy}$; $m/2\pi\vd$ equals the total density of edge
states $\qr_{\rm edge}$. The quantity $m\qt\vd$ that appears in
the higher dimensional operators is the 1D conductivity $\qs_{xx}$ of
$m$ channel edge states. Here, $2\qt\vd$ is the linear dimension which
sets the smallest wavelength for the $\widehat{Q}$ field variables and
$m/2$ is the (quantized) conductance ($g_m$) of the wire.

\vskip1cm
\scez\renewcommand{\theequation}{D\arabic{equation}}
\noindent
{\Large\bf Appendix D: Action for $Q$ and ${A}$ on multiple edges}
\addcontentsline{toc}{section}{Appendix D}

\vskip0.4cm
\noindent
The generalization of (\ref{afterlambdaS}) is given by
\bea
	S[Q,{A}] &=& \fr{\qb/2}{(2\pi)^2}\intd{^2 x d^2 x'}
	n(\vec x){B}\dagg(\vec{x})U_0(\vec{x}-\vec{x}')
	n(\vec x'){B}(\vec{x}')\nn\\
	&&+ \fr{i\qb}{4\pi}\left[
	\intdxx n(\vec x)\qe^{\mu\nu\qk}(A\eff_\mu)\dagg\prt_\nu A\eff_\qk
	+\sum_{a=1}^M s_a \oint_{C_a}\!\!\! dx 
	\left({A}_x\dagg {A}_0\eff
	-\fr{2\pi}{\qb}\tr \hat{A}_x Q \right)\right] \nn\\
	&& + \sum_{a=1}^M s_a \Stop^{(a)}[Q]
	+\fr{\pi}{4\qb\vd}\sum_{a=1}^M S_{\rm F}^{(a)}[Q] \nn\\ &&
	-\fr{\pi}{4\qb}\sum_{n\qa}\sum_{a=1}^M
	\int_{C_a}\!\!\! \fr{dk_x}{2\pi}\; \fr{1}{\veff(k_x)}
	\left|\vphantom{H^H}\tr \Ian Q(k_x)
	-\fr{\qb}{\pi}({A}_0\eff)^\qa_{-n}(k_x)\right|^2
\label{aaCoulomb}
	\\ &&
	+\fr{1}{8\qb\vd^2}\sum_{a\neq b}s_a s_b \oint_{C_a}\!\!\! dx
	\oint_{C_b}\!\!\! dx'\sum_{n\qa}
	[\tr\Ian Q-\fr{\qb}{\pi}({A}_0\eff)^\qa_{-n}](x) 
	\; \times \nn \\ && \times\; U_0(x,x')
	[\tr\Iamn Q-\fr{\qb}{\pi}({A}_0\eff)^\qa_n](x')
\label{anotbCoulomb}
\eea
where $U_0(x,x')$ denotes the full 2D Coulomb interaction.
All terms except those quadratic in $Q$ 
arise by the obvious replacements
$m\!\naar\! n(\vec x)$ and $m\oint \!\naar\! \sum_a s_a\oint_{C_a}$
in (\ref{afterlambdaS}).
The terms quadratic in $Q$ can be understood as follows.
In the generalized form of (\ref{afterWrot}), the quadratic term in the
plasmon field is given by 
\be
	-\fr{\qb}{2}\intd{^2 xd^2 x'}\ql(x)\dagg U_0\inv(x-x')\ql(x')
	+\fr{m}{2\pi\vd}\sum_a\oint_{C_a}\!\!dx\; \ql\dagg\ql,
\ee 
indicating that
the propagator for $\ql$ 
between two points on the same edge will be very different
from the propagator between different edges.
In the former case the propagator is proportional to
$[U_0\inv\!+\!\fr{m}{2\pi\vd}]\inv$, which is exactly the form obtained by
combining the Finkelstein term with (\ref{aaCoulomb}).
In the latter case, the propagator is simply proportional to $U_0$.
Finally, the signs can be understood by noticing that the 
coupling of the plasmon field to $Q$ is proportional to
$\sum_a s_a\oint_{C_a}\!dx\; \tr\hat\ql Q$.

\end{document}